%% file: main.tex
\documentclass[conference]{IEEEtran}
\usepackage{hyperref}
\usepackage{tikz}
\usepackage{amsmath}
\usepackage{filecontents}
\usepackage{graphicx}
\usepackage{multirow}
\usepackage{bigstrut}
\usepackage{upgreek}
\usepackage{array}
\usepackage{makecell}
\usepackage{xcolor}
\usepackage{soul}
\usepackage{booktabs}
\usepackage{tabularx}
\usepackage{amsmath}
\usepackage{float}
\usepackage{subcaption}
\usepackage{longtable}
\usepackage{times}
\usepackage[english]{babel}
\usepackage{siunitx}
\usepackage{amssymb}
\usepackage{array}
\usepackage{multicol}
\usepackage{etoolbox}
\usepackage{dblfloatfix}
\usepackage{fixltx2e}
\usepackage{adjustbox}
\usepackage{calc}
\usepackage{algorithm}
\usepackage{colortbl}
\usepackage[noend]{algpseudocode}

\newcolumntype{Y}{>{\centering\arraybackslash}X}

\usepackage{todonotes}

\usepackage{color}
\newcommand{\cmmnt}[1]{}

\newcommand{\etal}{\emph{et al. }}

\def \review #1{\color{black}#1\color{black}}

\begin{document}
%-------------------------------------------------------------------------------

\title{Evaluating the Information Security Awareness of Smartphone Users}

%\author{{\rm The Anonymous Authors}}
\author{Ron Bitton, Kobi Boymgold, Rami Puzis, Asaf Shabtai \\ Department of Software and Information Systems Engineering \\
Ben-Gurion University of the Negev \\
Beer-Sheva, 8410501, Israel}

\IEEEoverridecommandlockouts
\makeatletter\def\@IEEEpubidpullup{9\baselineskip}\makeatother
%\IEEEpubid{\parbox{\columnwidth}{
%    Network and Distributed Systems Security (NDSS) Symposium 2020\\
%    23-26 February 2018, San Diego, CA, USA\\
%    ISBN 1-1891562-49-5\\
%    https://dx.doi.org/10.14722/yyy.2020.23xxx
%    www.ndss-symposium.org
%}
%\hspace{\columnsep}\makebox[\columnwidth]{}}

\maketitle

%-------------------------------------------------------------------------------
\begin{abstract}
%-------------------------------------------------------------------------------
Information security awareness (ISA) is a practice focused on the set of skills, which help a user successfully mitigate a social engineering attack. 
Previous studies have presented various methods for evaluating the ISA of both PC and mobile users. 
These methods rely primarily on subjective data sources such as interviews, surveys, and questionnaires that are influenced by human interpretation and sincerity. 
Furthermore, previous methods for evaluating ISA did not address the differences between classes of social engineering attacks.
In this paper, we present a novel framework designed for evaluating the ISA of smartphone users to specific social engineering attack classes. 
In addition to questionnaires, the proposed framework utilizes objective data sources: a mobile agent and a network traffic monitor; both of which are used to analyze the actual behavior of users.
We empirically evaluated the ISA scores assessed from the three data sources (namely, the questionnaires, mobile agent, and network traffic monitor) by conducting a long-term user study involving 162 smartphone users.  
All participants were exposed to four different security challenges that resemble real-life social engineering attacks. 
These challenges were used to assess the ability of the proposed framework to derive a relevant ISA score. 
The results of our experiment show that: 
(1) the self-reported behavior of the users differs significantly from their actual behavior; 
and (2) ISA scores derived from data collected by the mobile agent or the network traffic monitor are highly correlated with the users' success in mitigating social engineering attacks.
\end{abstract}

\input{introduction.tex}

\input{background.tex}

\input{relatedwork.tex}

\input{method.tex}

\input{evaluation.tex}

\input{discussion.tex}

\input{conclusion.tex}

\bibliographystyle{plain}
\bibliography{main.bib}

\onecolumn
\section*{Appendices}
\subsection{\label{sec:behavior_qesutionnaire} The Security Questionnaire}
\input{Resources/Behavior_Questionnaire.tex}

\subsection{\label{sec:taxonomy_table} The Complete List of Security Awareness Criteria and the Indicators for Measuring Them}
\begin{center}
\input{Resources/Taxonomy_table_with_measurements.tex}
\end{center}

%%%%%%%%%%%%%%%%%%%%%%%%%%%%%%%%%%%%%%%%%%%%%%%%%%%%%%%%%%%%%%%%%%%%%%%%%%%%%%%%
\end{document}

%% file: introduction.tex
\section{\label{sec:intro} Introduction}
% **************** SE and ISA *****************
Social engineering (SE) is the process of exploiting human vulnerabilities in order to penetrate to a computerized system.
This is usually accomplished by manipulating a user to perform actions such as installing software, granting privileges, or revealing confidential information.
According to the Internet Security Threat Report (ISTR) published by Symantec in 2018 \cite{istr2018}, the chance that SE methods will be selected as an attack vector is greater than any other advanced technical exploit.
Hence, SE attacks pose a significant security threat to society. 

Social engineering attacks are not limited to PC users.
In fact, in the past five years the number of SE attacks targeting mobile devices users is rapidly increasing~\cite{istr2018}.
According to Avast and Mcafee mobile threat reports (2019) the majority of attacks on mobile devices exploit human factor vulnerabilities (such as downloading applications from unknown sources, clicking on suspicious links, granting unnecessary permission).
Furthermore, according to Checkpoint's security report, every organization had suffered a mobile malware attack during 2018.\footnote{https://www.checkpoint.com/downloads/product-related/report/2018-security-report.pdf}
When SE is the most popular way for spreading malware in mobile platforms~\cite{istr2018}, SE attacks targeting mobile users become a major concern to both individuals and organizations.

Information security awareness (ISA) is a practice focused on the set of skills which help a user to successfully mitigate SE attacks.
Since most SE threats are not covered by technological countermeasures, ISA is the most effective method to prevent SE attacks.
Thus, assessing the ISA level of employees and thereby identifying those who are more vulnerable to SE attacks is extremely important for organizations. 
Early identification of such employees can be used to improve the security level of the organization, for example, by updating their security policies \cite{reason2016managing} and implementing personalized security awareness programs \cite{aytes2003research, kabay2002using, thomson1998information}.

Previous studies have presented various methods for evaluating and increasing the ISA of PC users \cite{thomson1998information,kabay2002using,aytes2003research,kruger2006prototype,kruger2006framework,reason2016managing}.
However, the challenges in evaluating the security awareness of smartphone users are different from PC users for several reasons.
First, smartphones are equipped with unique/different functionalities that are less common in PCs (location services, motion sensors, SMS, application stores and etc.)
Second, smartphone users consume content and services mostly via apps, opposed to browsers in PCs.
Third, some security safeguards are less common on smartphones than on PCs (e.g., antivirus or IDS), while others are implemented and used differently (permissions, certificates handling, encryption storage, authentication mechanism).
Finally, the threats are different. 
According to the Internet security threat report (ISTR)~\cite{istr2018}, the top threat categories for smartphone users are fake apps, spyware, SMS messages, and permission abuse; in contrast PC users top threats are spear phishing, malicious websites, Trojans, and web-server exploits.
These main differences clearly indicate that different security awareness skills are required by a mobile user for operating safely mobile devices, compared to a PC.

Most current studies on ISA assessment depend excessively on the subjects' response to questionnaires~\cite{mccormac2016test} or surveys~\cite{albrechtsen2010improving}. 
These methods rely on the subjects' self-reported behavior, and thus, tend to be subjective and biased \cite{redmiles2018asking}. 
Moreover, these methods require the subjects' active involvement and collaboration, thereby consuming significant human resources. 

Other methods are based on measuring the momentary behavior of subjects during specific events (e.g., by monitoring their reaction while facing a simulated SE attacks)~
\cite{SET,jansson2013phishing,kumaraguru2009school,doi:10.1108/IMCS-11-2013-0083}.
These methods, however, tend to be sensitive to environmental/contextual factors and therefore cannot provide a reliable ISA score~\cite{conway2017qualitative}. 
For these reasons, previous studies have argued that in order to reliably assess the ISA, continuous measurement of the subjects' actual behavior must be considered~\cite{kruger2006framework,kruger2006prototype,redmiles2018asking}. 
Although such measurements were developed for the PC~\cite{forget2014security, wash2017can}, the criteria for a security aware smartphone user are different from those of a PC user~\cite{bitton2018taxonomy}. 
To the best of our knowledge, a comprehensive ISA assessment method, which encompasses the various aspects of mobile security and is based on continuous measurement of users' actual behavior has not yet been developed for mobile platforms. 

In this paper, we suggest a novel framework for assessing the ISA of smartphone users. 
The framework utilizes data collected and analyzed from three different data sources to derive/measure a set of  criteria for a security aware smartphone user: (1) a \emph{mobile device agent}, which is installed on the subjects' devices and measures the subjects' actual behavior (e.g., applications installed, security settings, connectivity management, and browsing history), (2) a \emph{network traffic monitor}, which analyzes the network traffic sent to/from the subject's devices, and (3) a \emph{security questionnaire}, specifically designed to assess the self-reported behavior of smartphone users.

The expert-based process suggested in~\cite{bitton2018taxonomy} is utilized for deriving an ISA model that is specific to an SE attack class.
We define a \emph{SE attack class} as the set of SE attacks that exploit similar human vulnerabilities (e.g., phishing, application-related, and MITM attacks).
Each model defines the contribution (i.e., weight) of each criterion to the overall ISA score of a specific SE attack class and can be applied for deriving the ISA score from the measured criteria.
The ISA score provides an indication of the ability of the user to mitigate a particular class of SE attack.

In order to evaluate the proposed framework we conducted an empirical experiment involving 162 smartphone users for a duration of seven to eight weeks each.
In addition to the data collected from the three data sources mentioned above, the participants were exposed to four different security challenges that resemble real-life SE attack scenarios: a phishing attack, an eavesdropping attack, a certificate manipulation attack, and a privilege escalation attack.

To validate the proposed framework, we tested the correlations between the derived ISA scores and the results of the challenges.
We also analyzed the correlation between ISA scores derived by the different data sources.
In addition, we systematically compared the data extracted by the mobile agent, network traffic monitor and challenges with the questionnaire results in order to examine the bias in the questionnaires compared to objective measurements within the context of information security.

The results of the experiment show that measurements of the subjects' actual behavior using the mobile agent/network traffic monitor produce ISA scores that  correlate with the subjects' success or failure in the challenges.
Our findings also question the reliability of assessing the ISA of smartphone users based on their self-reported behavior.  
In our experiment, subjects who reported to behave cautiously with respect to smartphone security failed in multiple social engineering challenges, while subjects who acknowledged their unsafe behavior successfully mitigated most of the challenges.

Moreover, we show that the ISA scores derived from the mobile agent are highly correlated with the ISA scores derived from the network traffic monitor. 
In contrast, the ISA scores derived from the questionnaire data source were found to have a very little correlation with the scores derived from the mobile agent and network traffic monitor data sources.

In summary, the contribution of the proposed research are two fold: 

\begin{itemize}
\item \textbf{A framework for evaluating the ISA of smartphone users.}
The first contribution of this research is a framework that can automatically detect users who are more vulnerable to social engineering attacks.
The proposed framework relies on actual user behavior as observed from multiple data sources, as opposed to commonly used subjective self-reported behavior methods which were found to be unreliable in some cases. 
Moreover, in contrast to previous works, which explore the ISA of smartphone users with respect to a \emph{specific} information security aspect (e.g., permissions, lock screen, etc.), the proposed method integrates \emph{multiple} mobile security aspects into a single ISA score per attack class.

\item \textbf{A comprehensive user study.}
The second contribution of this research is a comprehensive and very unique user study.
Within this study we monitored, using four different information sources (questionnaire, mobile agent, network traffic and challenges), the behaviour of 162 users while using their own smartphones, as opposed to previous works that either measure the momentary behavior of the subjects during specific events, or performed the assessment in a controlled environment, by using some dedicated device/platform.
This unique user study allow us to explore the bias in questionnaires compared to objective measurements with respect to multiple behaviours within the context of information security.
\end{itemize}

%% file: background.tex
\section{\label{sec:background} Background}
Kruger and Kearney developed a prototype for assessing the ISA in organizations \cite{kruger2006prototype}.
The authors suggest that stakeholders define \emph{focus areas}, i.e., primary technological issues related to the organization's information security, and explore them within the context of three psychological dimensions: \emph{knowledge}, \emph{attitude}, and \emph{behavior}.
Each dimension addresses a specific psychological aspect of the focus area and contains factors that assist in the quantitative measurement of the dimension.

Bitton \etal~\cite{bitton2018taxonomy} redefined and adjusted the prototype presented by Kruger and Kearney for the mobile platform.
The authors presented an extensive list of measurable indicators, also referred to as criteria, for a security aware smartphone user.
In addition, the authors introduced the concept of deriving the ISA level of a user with respect to a given class of SE attacks i.e., attacks that exploit similar human vulnerabilities.
The motivation behind this concept is that the skills required to mitigate one class of attacks may differ from the skills required to mitigate another class of attacks.

\begin{figure}[t]
\centering
    \fbox{\includegraphics[width=0.48\textwidth]{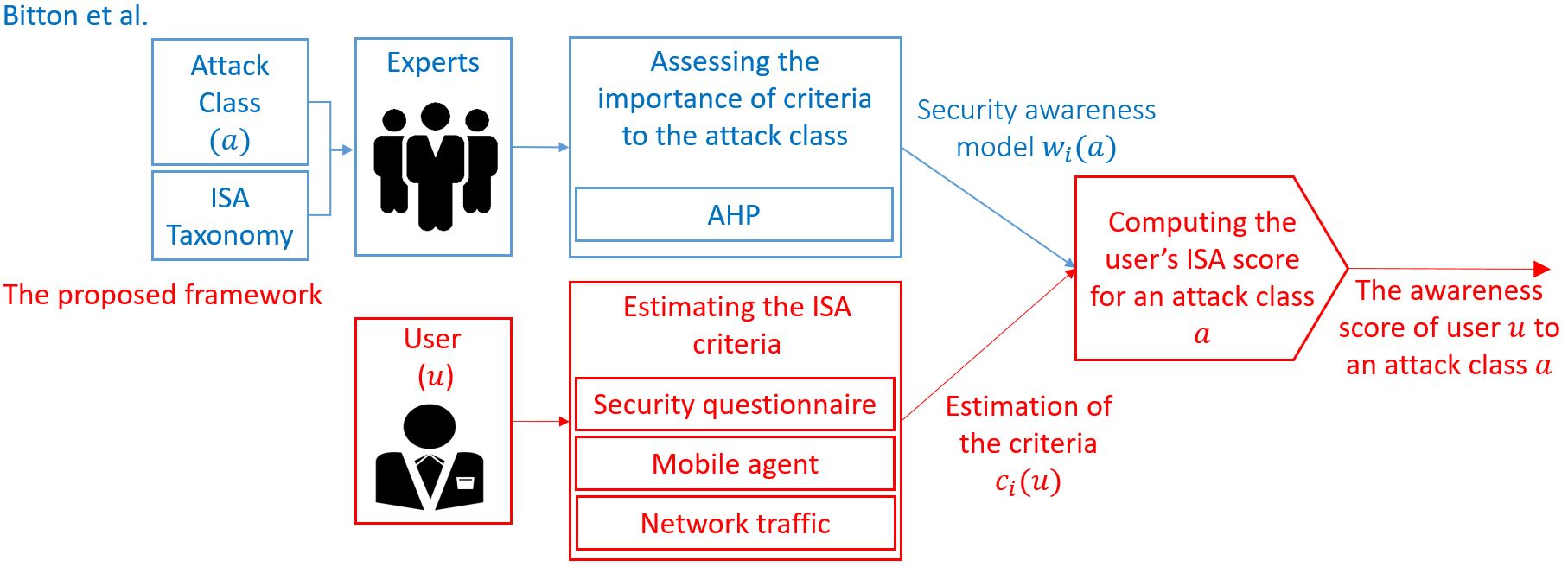}}
    \caption{\label{fig:score-calc} The process of computing the security awareness score of a user for an attack class}  
\end{figure}

To assess the ISA of users for a specific class of attacks, Bitton \etal suggested the expert-based process presented in Figure \ref{fig:score-calc} (in blue). 
As can be seen, the inputs to this process are: 
(1) an attack class (denoted by $a$),  and (2) a list of criteria for a security aware smartphone user (listed in the ISA taxonomy~\cite{bitton2018taxonomy}).
Security experts utilize the analytic hierarchy process (AHP)~\cite{saaty2008decision} to efficiently estimate the contribution of each criterion to the ability of users to mitigate SE attacks of particular classes.
The outcome of this process is a \emph{security awareness model}, which is a set of weights $w_{(i)} (a)$ that reflect the contribution of each criterion $i$ to the ability of users to mitigate attack class $a$.

With the assistance of 17 security experts from both academia and industry, Bitton \etal applied the process and derived security awareness models three attack classes: \emph{Man-in-the-middle attacks} such as session/signal hijacking, SSL strip, eavesdropping, and rogue access point;
\emph{Application related attacks} such as application phishing, malicious notifications (e.g., error messages), in app pop-ups, malicious advertisements, clicking fraud, Trojan applications, and rootkits; and \emph{Phishing attacks} such as social network fraud (e.g., fake links, friend/game requests), phishing emails, websites, URL, SMS, IRC, forums.

In this paper we extend the work presented in \cite{bitton2018taxonomy} by introducing a comprehensive framework for measuring the criteria for a security aware smartphone user and computing the ISA score of smartphone users to different classes of SE attacks.
The process of computing the ISA score of a smartphone user (denoted as $u$) for a specific class of SE attack (denoted as $a$) is presented in Figure \ref{fig:score-calc} (in red).
As can be seen, there are two inputs to the computation process: (1) the security awareness model for the particular class of attacks as defined by Bitton \etal, and (2) an estimation of all criteria for a particular user (i.e., the set of measured values for each criterion $i$ for user $u$), as measured by the proposed framework.

%% file: relatedwork.tex
\section{\label{sec:related-work} Related Work}
The domain of ISA has been researched extensively in the past.
In this section, we describe previously suggested methods for assessing the ISA of mobile users.
Specifically, we present related work that used questionnaires as the data source (Section \ref{sub-sec:related-work-questionnaires}), research that used software agents as a passive measurement data source (Section \ref{sub-sec:related-work-passive-measurements}), and related work that applied active challenges, either for evaluating a proposed method of ISA measurement or as a tool for training users (Section \ref{sub-sec:related-work-attack-scenarios}).
In addition, we summarized all relevant papers in the domain of ISA assessment in Table \ref{tab:relatex_work_table}.
\input{Resources/related_work_table.tex}

\subsection{\label{sub-sec:related-work-questionnaires} The Assessment of ISA Using Questionnaires}
\textbf{PC users.} A popular method for assessing the ISA of PC users is the Human Aspects of Information Security Questionnaire (HAIS-Q) \cite{parsons2014determining}, which aims to measure the knowledge, attitude, and self-reported behavior of employees regarding information security.
The external validity of HAIS-Q was examined in previous studies, by using the test-retest method \cite{mccormac2016test} and conducting phishing tests \cite{parsons2017human}.
Onarlioglu \etal studied the resilience of PC users to phishing attacks by asking participants to rank the potential risk resulting from certain scenarios \cite{onarlioglu2012insights}.

While Parsons \etal~\cite{parsons2017human, onarlioglu2012insights} focused on the PC domain and conducted an evaluation involving one attack vector (i.e., phishing tests), in this study we focus on smartphone users and evaluate the proposed method on multiple SE attacks.
In addition, the evaluation presented in these works was performed in a controlled environment (by asking subjects to assess the safety of clicking links in a series of simulated phishing emails), which may not reflect the subjects' behavior in a real attack scenario~\cite{brewer2000research}.
In this study, the various attack simulations are executed in the natural environment of the users, which totally reflects the subject's behaviour in a real attack scenario.
Furthermore, in contrast to the proposed research these studies do not validates the self reported behavior of the users using objective data sources (e.g., an agent which installed in the user's PC and measure the actual behaviour 
of the user)

\textbf{Mobile users.} Androulidakis and Kandus studied the security awareness of smartphone users through questionnaires \cite{androulidakis2011mobile, androulidakis2011survey, androulidakis2011bluetooth}.
Based on their questionnaires, Esmaeli~\cite{esmaeili2014assessment} suggested a model for assessing factors that affected information security behavior in smartphone networks.
Mylonas \etal studied the ISA of smartphone users in the context of downloading applications from official application repositories (e.g., Google Play, and Apple's App Store)~\cite{mylonas2013delegate}.
Gkioulos \etal presented an analysis of a security questionnaire, designed specifically for assessing the ISA of smartphone users~\cite{gkioulos2017security, gkioulos2017user}.
Their results indicate that for several security topics, the security competence level of users does not affect the subjects’ self-reported behavior. 
Each of these works contributed significantly to the design of the security questionnaire proposed in this study.

Nevertheless, previous works have three main limitations.
First, prior works explore the ISA of smartphone users with respect to \emph{specific} information security aspect. 
As summarized on Table~\ref{tab:relatex_work_table} many important aspects of smartphone security awareness (such application usage, browsing, and account security) are not covered by these works.
In addition, the integration of different information security aspects into a single comprehensive and complete framework for assessing the ISA of users was not presented in prior works.
Second, these works do not validates the self reported behaviour of users using objective data sources (such as a mobile agent and network traffic monitoring).
In this work, we address this gap by developing three orthogonal methods for assessing the ISA of users, a security questionnaire, mobile agent and network traffic monitor.
We also compared the results of the three assessment methods and provides meaningful insights with regards to their reliability and accuracy.
Third, previous studies was mostly evaluated using very limited set of attack vectors (typically different levels of phishing emails), which in most cases was performed in a controlled environment (which may not reflect the subjects' behavior in a real attack scenario).
In this study we evaluate the framework by measuring his/her reaction when facing multiple real-life attack scenarios.

\subsection{\label{sub-sec:related-work-passive-measurements} The Assessment of ISA Using Passive Measurements}
\textbf{PC users.} Forget \etal introduced the Security Behavior Observatory (SBO), a client-server architecture designed for long-term monitoring of client machines running the Windows operating system~\cite{forget2014security}.
In follow-up research they utilized the SBO to validate the self-reported behavior of users, assessed through interviews~\cite{forget2016or}.
Wash \etal also presented a data collection system to validate the self-reported behavior of PC users, assessed through surveys~\cite{wash2017can}.

Redmiles \etal compared real measurement data to survey results with respect to a specific user behavior: software updating~\cite{redmiles2018asking}.
Their results indicate that for software updating the self-reported data varies consistently and systematically with measured data.

Although these studies included objective measurements and continuous monitoring of the subjects' actual behavior, they focused on very specific user behaviors.
In addition, they centered on PCs rather than smartphones.
Hence, these studies did not cover technological aspect that are relevant for smartphone ISA, such as location-based services, motion sensors, microphones, permissions and etc.
Furthermore, these studies was never evaluated using real-life attack scenarios.
Nonetheless, their practical results support the findings presented in this paper, namely the reports of users regarding their behavior are unreliable.

\textbf{Mobile users.} The self-reported behavior of users tend to be biased~\cite{ajzen1991theory, krumpal2013determinants}.
Consequently, ISA assessment based on users' self-reported behavior tends to be inaccurate. 
Because of this, researchers have developed objective measurement tools for measuring the actual behavior of users.
Egelman \etal developed and examined the validity of SeBIS~\cite{egelman2015scaling}, a security questionnaire intended to measure the attitude of users towards security-related topics~\cite{egelman2016behavior}.
In their study, Egelman \etal examined the validity of their questionnaire using three distinct experiments: (1) by requesting the subjects to suggest a strong password and identify phishing URLs, (2) by examining the operating system version of the subject’s personal computer two weeks after a release of an update, and (3) by checking for the existence of a secure lock screen on the subject's personal smartphone. 

The authors of~\cite{van2013modifying} also studied the ISA of smartphone users in the context of lock screen security (e.g., pattern, textual password). 
In their experiment, the authors provided smartphones to the subjects with unlimited data, SMS, and mobile-to-mobile use in exchange for being monitored while participating in the study. 
During the course of the experiment, the subjects also completed several surveys to assess the factors that influenced the subjects' decision to use a lock screen code or share this code with acquaintances. 

Wijesekera \etal studied the perception of smartphone users regarding permission~\cite{wijesekera2015android}.
In their experiment, the authors collected contextual information about the state of the device when permissions are requested.
At the end of the experiment, in a concluding survey, they asked the subjects about the permissions granted and denied as well as their decision making in accordance to specific contexts.

These works have four main limitations.
First, the above methods explore the ISA of smartphone users with respect to \emph{specific} information security aspect.
For instance, Egelman \etal~\cite{egelman2016behavior} focused solely on password usage, OS version and lock screen; Van \etal~\cite{van2013modifying} was also focused solely on lock screen security; and Wijesekera \etal~\cite{wijesekera2015android} focused solely on application permissions. 
As summarized in Table~\ref{tab:relatex_work_table}, many important aspects of smartphone security awareness was not covered by these works.
In addition, the integration of different information security aspects into a single comprehensive and complete framework for assessing the ISA of users was not presented in prior works.

In this work, we address this gap by presenting a comprehensive and complete framework for assessing the ISA of smartphone users. 
Within this framework, we develop multiple sensors that collects device's security attributes and monitor the behaviour of the user, with respect to multiple information security aspect (the complete list of sensors is summarized in Table~\ref{tab:agent-sensors-list} in Appendix~\ref{sec:sensor_list_table}).
In addition, based on an expert-based procedure, we integrate the various mobile security aspects into a single ISA score per attack class.

Second, previous studies are either measuring the behavior of the subjects within a specific context (e.g., updating a specific program) and in a controlled environment~\cite{egelman2015scaling}, or provide additional smartphone to the participant for the experiment period.
Thus, may not reflect the actual behaviour of the user when using his own smartphone on his natural environment.

Third, previous studies were not evaluated using 
real-life social engineering attacks.
Hence, it is hard to estimate their impact to the information security of an organization in a real-life deployment.
In this study we evaluate the framework by measuring his/her reaction when facing multiple real-life attack scenarios.

Finally, to the best of our knowledge, the assessment of user's ISA through network traffic was never presented in prior research.
The contribution of this kind of assessment is numerous, since it provides a completely passive method for assessing user's ISA (i.e., without any interaction with the subject or subject's device).

\subsection{\label{sub-sec:related-work-attack-scenarios} The Assessment of ISA Using Attack Scenarios}
Previous studies have developed methodologies for designing experiments that include security challenges.  
Dodge \etal~\cite{dodge2007phishing} outlined the aspects to consider when designing a phishing experiment in which the efficacy of a security awareness program is evaluated. 
The authors of~\cite{jagatic2007social} utilized publicly available information on social networks to automatically execute spear phishing challenges. 
Their results show that the ability of subjects to mitigate phishing attacks decreases dramatically when the origin of the message is one of the subjects' acquaintances. 

Security challenges (i.e., security exercises) have also been used as a method for assessing the ISA of users. 
The authors of~\cite{doi:10.1108/IMCS-11-2013-0083} utilized surveys, interviews, and phishing challenges to explore the factors that affect the ability of subjects to mitigate phishing attacks, as well as the correlation between self-reported and observed behavior. 
Although their study was not targeted at the mobile domain, their results support our conclusion that the self-reported behavior of subjects does not necessarily reflect their behavior when faced with a real attack scenario~\cite{brewer2000research}. 

Kumaraguru \etal introduced PhishGuru, an embedded training system that teaches subjects how to mitigate phishing attacks~\cite{kumaraguru2007protecting}.
In subsequent research, the authors performed a real-world evaluation of PhishGuru with 515 subjects for a period of 28 days. 
The results of their evaluation show that PhishGuru is effective~\cite{kumaraguru2009school}. 
The author of~\cite{jansson2013phishing}, implemented a phishing experiment based on the methodology presented in~\cite{dodge2007phishing}.
Their results show that phishing exercises can improve the subjects' ISA with regard to phishing.

While SE challenges have been used in the past for assessing the ISA of users, the vast majority of them were focused solely on phishing attacks, and were not targeted at the mobile platform. 
In this study we implement four different challenges (Namely, phishing challenge, SPAM, Permission abuse and certificate manipulation). 

%% file: Resources/related_work_table.tex
\begin{table*}[t]
\scriptsize
\begin{tabularx}{\linewidth}{| 
>{\centering}m{.1065\linewidth} |
>{\centering}m{.08\linewidth} |
>{\centering}p{.0155\linewidth}
>{\centering}p{.0155\linewidth}
>{\centering}p{.0155\linewidth}
>{\centering}p{.0155\linewidth}
>{\centering}p{.0155\linewidth}
>{\centering}p{.0155\linewidth}
>{\centering}p{.0155\linewidth}
>{\centering}p{.0155\linewidth}
>{\centering}p{.0155\linewidth} | 
>{\centering}m{.202\linewidth} | 
>{\centering\arraybackslash}m{.165\linewidth} |}

\hline
\multirow{2}{*}{\vspace{-3cm} Paper} &
\multirow{2}{*}{\vspace{-3cm} Domain} &
\multicolumn{9}{c|}{Focus Areas} &
\multirow{2}{*}{\vspace{-3cm} Data Source} &
\multirow{2}{*}{{\parbox{\linewidth}{\centering\vspace{1cm} Evaluated Using Attack Scenarios}}}\\
\cline{3-11}

    & & 
   \raggedright \rotatebox{90}{Application Installation} &
   \raggedright \rotatebox{90}{Application Handling} &
   \raggedright \rotatebox{90}{Browser} &
   \raggedright \rotatebox{90}{Account} &
   \rotatebox{90}{Virtual Communication } &
   \rotatebox{90}{Operating System} &
   \rotatebox{90}{Security Systems} &
   \rotatebox{90}{Network} &
   \rotatebox{90}{Physical Channels} & & \\
 \hline
 
\cite{mccormac2016test},\cite{parsons2014determining},\cite{parsons2017human} & PC, Mobile & No & No & Yes & Yes & Yes & No  & No & Yes & Yes & Questionnaire & Controlled Environment  \\ 

\cite{androulidakis2011mobile},\cite{androulidakis2011survey}, \cite{androulidakis2011bluetooth} & Mobile & Yes & No & No & Yes & Yes & Yes & Yes & Yes & Yes & Questionnaire & No\\

\cite{esmaeili2014assessment} & Mobile & Yes & No & No & No & No & No & Yes & No & Yes & Questionnaire & No\\

\cite{mylonas2013delegate} & Mobile & Yes & Yes & No & No & No & Yes & Yes & No & Yes & Questionnaire & No\\

\cite{gkioulos2017security},\cite{gkioulos2017user}  & Mobile & Yes & Yes & No & Yes & No & Yes  & No & Yes & Yes & Questionnaire & No \\

\cite{egelman2016behavior},\cite{egelman2015scaling} & PC, Mobile & No & No & No & Yes & Yes & Yes & Yes & No & No & Questionnaire, passive measurements & No \\

\cite{onarlioglu2012insights} & PC & No & No & Yes & No & Yes & No & No & No & No & Laboratory tests & No \\
%\hline
% TODO: the following is not actually an ISA assessment study, but a study of the permission dialogues in general
%\cite{wijesekera2015android} & Mobile & Yes & Yes & No & No & No & No & No & No & No & No & Monitoring permission dialogues & No \\
%\hline
  
  \cite{wash2017can} & PC & No & Yes & Yes & Yes & No & Yes & No & No & No & Survey, PC agent & No \\
  
  \cite{forget2016or},\cite{forget2014security} & PC & Yes & Yes & No & No & No & Yes & Yes & No & No & Interviews, PC agent & No \\
  
%  \cite{kumaraguru2009school},\cite{dodge2007phishing},\cite{jagatic2007social} & PC & No & No & No & No & Yes & No & No & No & No & No & Simulated phishing emails & Yes \\

\cite{doi:10.1108/IMCS-11-2013-0083} & PC & No & No & No & No & Yes & No & No & No & No & Surveys, and interviews & Yes \\

 \cite{van2013modifying} & Mobile & No & No & No & No & No & No & No & No & No & Mobile agent & No \\
%\hline
 
 %cite & domain & AI & AH & B & A & VC & OS & DP & SS & N & PC & Measurement tool & evaluation \\
% cite & domain & AI & AH & B & A & VC & OS & DP & SS & N & PC & Measurement tool & evaluation \\
\hline
\rowcolor[gray]{0.9}
The proposed framework & Mobile & Yes & Yes & Yes & Yes & Yes & Yes & Yes & Yes & Yes & Questionnaire \\ Mobile agent \\ Network traffic analysis & Yes\\
 \hline
\noalign{\hrule height 1pt}
\end{tabularx}
\caption{Summary of existing methods for assessing the ISA of users: as can be noticed, previous works are mainly rely on subjective data sources, are not validated using attack scenarios in an uncontrolled environment (with the exception of \cite{doi:10.1108/IMCS-11-2013-0083}, which used a single phishing experiment), and are not encompass all relevant focus areas (in most cases only small set of security topics were considered in each focus area.)}\label{tab:relatex_work_table}
\end{table*}

%% file: method.tex
\section{\label{sec:framework}The Proposed Framework}
The proposed framework encompasses three types of data sources: a mobile device agent, which is installed on the subjects' smartphones and measures their actual behavior; a network traffic monitor which extracts a set of behavioral properties from the network traffic transmitted to/from the subjects' smartphones; and a security questionnaire designed to measure the ISA of smartphone users.
Each data source can be used separately to assess the ISA of a user; thus, the assessment process will result in three ISA scores for each user and attack class.

\subsection{\label{subsec:mobile-agent} Mobile Device Agent}
The first data source is implemented as a software application (agent) installed on the subjects' smartphones.

Using a variety of dedicated software modules (i.e., sensors), the agent continuously monitors the actual behavior of the subjects while operating their smartphone.
Thus, it provides a complete, objective, and continuous record of the subjects' actual behavior.

These dedicated sensors are used to estimate the ISA criteria presented in \cite{bitton2018taxonomy}, which can later be used for assessing the ISA of a given subject for a specific class of SE attacks.
The complete list of ISA criteria, as well as their indicators from the mobile agent is presented in Appendix~\ref{sec:taxonomy_table}.

The sensors implemented  can be categorized into four groups: connectivity, applications, content, and device.
The complete list of sensors is summarized in Table~ \ref{tab:agent-sensors-list}.
\input{Resources/Sensors_list.tex}

\subsubsection{Connectivity}
The connectivity of a device can pose a threat to the user.
For instance, the Bluetooth protocol has already been used many times as an attack vector~\cite{hypponen2006malware, androulidakis2011bluetooth, peng2014smartphone, la2013survey}. 
Similarly, untrusted Wi-Fi networks are fertile ground for network attacks such as eavesdropping and man-in-the-middle (MITM)~\cite{la2013survey}.

Within the proposed framework, we developed software modules, which are used to monitor the behavior of the user while operating the device's connectivity sensors (e.g., Wi-Fi, Bluetooth, GPS, and NFC), and collect statistical information regarding the incoming/outgoing network traffic of each application.

Specifically, we utilize the data extracted from these sensors to evaluate various criteria related to \emph{communication channels}.
For instance, to identify a potential risk of eavesdropping and content hijacking, we explore connections to unencrypted Wi-Fi networks to detect files and private data transmitted as plain-text over these insecure channels (N1, N2, N4 in Appendix~\ref{sec:taxonomy_table}).
We also examine whether the user has privacy concerns with respect to connectivity, by observing whether the user manages the connectivity sensors cautiously, i.e., disables a connectivity when it is not in use (PC1 in Appendix~\ref{sec:taxonomy_table}).

\subsubsection{Applications}
The applications installed on the smartphone have the ability to access the user's sensitive data, as well as to collect sensors' data in real-time (e.g., camera, microphone, GPS, etc.); thus, malicious applications can pose a threat to the user.

The application sensors are used for collecting indicators of careless user behavior, with respect to installation and management of applications. 
As applications from dubious sources are more likely to be malicious~\cite{mylonas2013delegate, felt2011survey, jiang2012dissecting}, the source of the application is a very important indicator for the ISA of a user with respect to applications.
For this reason, we search for such installation and use third party services (such as VirusTotal~\cite{virustotal}) to identify malicious applications (AI1 in Appendix~\ref{sec:taxonomy_table}).
We also search for applications that require root permission, since careless user behavior with rooted devices may permit applications to bypass the application's sandbox mechanism~\cite{peng2014smartphone, damopoulos2011isam} (AI4 in Appendix~\ref{sec:taxonomy_table}).

Since the permission mechanism can be easily exploited through SE~\cite{vidas2011all, kelley2012conundrum, mylonas2014assessing, haris2014privacy}, we examine the permissions requested by applications with a low rating or low number of downloads in order to identify sensitive permissions granted to suspicious applications (AI2,AI3 in Appendix~\ref{sec:taxonomy_table}).

The applications installed on a subject's device can also indicate reckless behavior.
Thus, we check for the existence of anti-virus software, as well as password management, backup and VPN services, since their existence corresponds to cautious user behavior (SS2, N3, and A3 in Appendix~\ref{sec:taxonomy_table}). 
Moreover, we continuously monitor updates and deletions of applications to validate that the user manages his/her applications carefully (AH1, AH3 in Appendix~\ref{sec:taxonomy_table}).   

\begin{figure}[t]
\centering
    \includegraphics[width=0.49\textwidth]{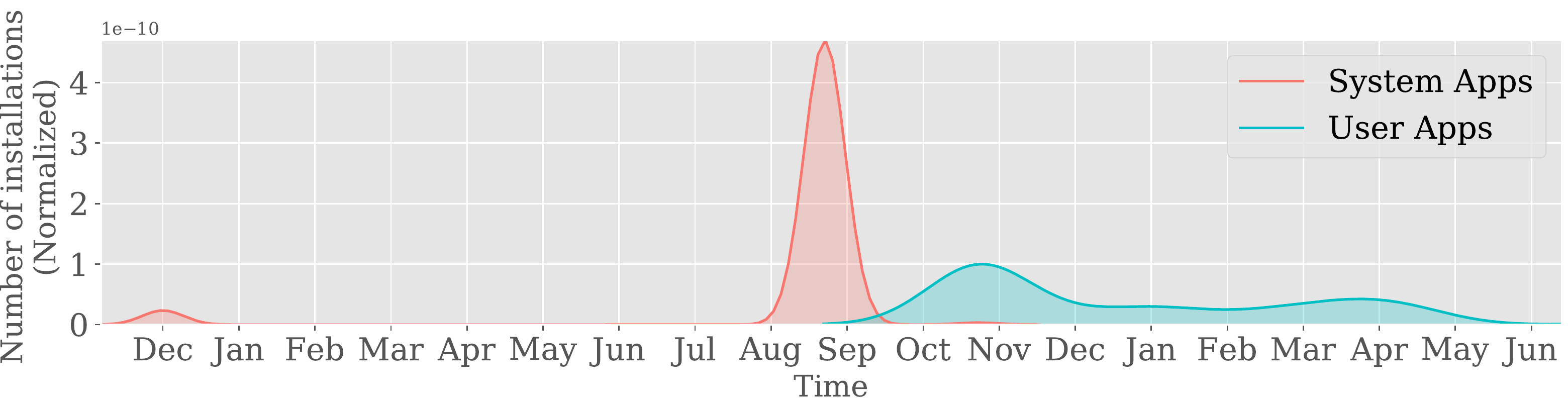}
    \caption{\label{fig:Installation-times} A distribution of the installation times of system and user applications for a specific sampled smartphone user.}
\end{figure}

A challenging task with regard to the data mining of the application sensors was identifying the specific applications installed by the subjects themselves.
As different vendors include different system applications on their devices,
these applications, which typically require many sensitive and dangerous permissions, are mixed in with the users' applications.
Nonetheless, we noticed that system applications are usually installed in a burst that precedes the installation of applications by a user, sometimes many months before the first user-installed application.
In contrast, user applications are installed gradually along a long period of time. 
Figure~\ref{fig:Installation-times} presents a sample distribution of installation times for both user and system applications for a sampled smartphone user.
We can clearly see the burst of system application installations that precedes the installations of the first user applications.

In addition, it can also be noticed that user applications are installed over a period of nine months (with a slight increase in the number of user applications installed at the beginning of this period).

For this reason, in order to filter out the system applications, we first sort the subjects' application lists by the application's installation time; then, we eliminate large installation bursts that appeared early in the installation timeline.

\subsubsection{Content}
Smartphones have changed the way we consume data and communicate with each other~\cite{karikoski2013contextual}.
In 2018, 52.2\% of all website traffic was generated by mobile phones.\footnote{https://www.statista.com/statistics/241462/global-mobile-phone-website-traffic-share/}
Furthermore, 55\% of all the emails opened, were opened through mobile devices,\footnote{https://returnpath.com/downloads/email-client-experience/}
thus exposing smartphone users to additional threats such as spam, phishing, and account hijacking.

The content sensors monitor statistical information regarding the browsing history, email messages, and SMS messages.
Specifically, we use the browsing data to identify malicious domains, e.g., by using VirusTotal~\cite{virustotal}); and to detect expired and self-signed SSL certificates, e.g., by using OpenSSL \cite{openssl}
(B1, B5 in Appendix~\ref{sec:taxonomy_table}). We utilize the default email API, as well as the Gmail API~\cite{gmailapi} to analyze the spam mailbox and detect the opening of spam emails (VC1 in Appendix~\ref{sec:taxonomy_table}).
We monitor malicious links sent via SMS messages (VC1 in Appendix~\ref{sec:taxonomy_table}); by crosschecking the SMS information with the browsing history, we identify connections made to malicious domains that were initiated using SMS messages (VC2 in Appendix~\ref{sec:taxonomy_table}).
In addition, we identify the use of two-factor authentication services by searching for SMS messages with security codes (A2 in Appendix~\ref{sec:taxonomy_table}).

\subsubsection{Device} 
The device's state and settings are very important aspects of smartphone security awareness.
One of the most popular attack vectors in smartphones is the exploitation of known vulnerabilities.
Thus, maintaining an up to date operating system is an extremely important countermeasures against cyber attacks on smartphones.

Within the framework, we develop sensors that collects the device's security attributes.
For instance, we examine whether the device's operating system is up to date (OS1 in Appendix~\ref{sec:taxonomy_table}).
We explore whether the user protects his/her data from unwanted physical access, by validating lock screen, PIN code and pattern use (SS5 in Appendix~\ref{sec:taxonomy_table}).
We also test whether the device is rooted (OS2 in Appendix~\ref{sec:taxonomy_table}).

\subsection{\label{subsec:gateway-module} Network Traffic Monitor}
The second data source is a network traffic monitor, which extracts a set of behavioral attributes from the Internet traffic of the subjects.
These attributes are later used to evaluate the ISA criteria that serve as the input of the process of assessing a subject's ISA for a specific SE attack class.

The primary advantage of this data source is rooted in its passiveness.
In contrast to questionnaires and mobile agents, which require some level of collaboration and interaction with the subject, network traffic monitoring is passive, taking place with no user interaction or involvement whatsoever.

The processes for extracting the behavioral attributes using a network traffic monitor can be divided into four categories: domain, application level protocols, deep packet inspection (DPI), and contextual analysis.

\subsubsection{Domain}
The domains exchanged within application layer protocols, such as the HTTP host and SSL server name, can provide indications regarding the user’s behavior. 
For instance, recurrent TCP sessions to anti-virus domain  can indicate that the user has an anti-virus application.

In order to identify indicators of cautious/reckless behavior we extract the domains from the raw network traffic of users.
Using third party services (such as Web of Trust (WoT) \cite{wot}, VirusTotal \cite{virustotal}, BitDefender, Dr. Web, and Websense ThreatSeeker), we classify the domains into two types of categories:
(1) categories that are related to information security (e.g., scam, spam, ads/pop-ups, malware or viruses, privacy risks, etc.); and (2) categories that are related to the content of the session (e.g., social network, email, news, forum, etc.)
This classification allows us to identify reckless behavior such as connecting to malicious sites (B1 in Appendix~\ref{sec:taxonomy_table}).

The domain's data was also used to identify the communication with common analytic services (such as \emph{ad.doubleclick.net}), which provide indications about pop-up and advertisement clicks (AH2 in Appendix~\ref{sec:taxonomy_table}). 
In addition, we identify recurrent connections to domains that are associated with security applications (e.g., anti-virus, password management services, etc.) to detect the presence of security countermeasures (A3, SS2, SS3 in Appendix~\ref{sec:taxonomy_table}).

\subsubsection{Application level protocols} 
In this category we focus on two common application level protocols: HTTP and TLS.
HTTP is an application level protocol widely used for hypertext information systems (e.g., Web applications).
The header of an HTTP packet includes general information about the smartphone.
Specifically, the user-agent field specifies the device model and OS version.
We use this information to understand whether the user is constantly updates his/her device's operating system(OS1 in Appendix~\ref{sec:taxonomy_table}).

TLS (and its predecessor, SSL) is an application layer protocol used to provide secured communication over untrusted networks. 
TLS provides server authenticity using digital certificates which are transmitted from the server to the client during the handshake.
TLS security is rooted in users' trust of the certificate authorities that issue these certificates. 
Accepting an untrusted certificate (e.g., self-signed, expired, invalid signature, etc.) is considered risky behavior, since it exposes the user to various attack vectors, such as phishing and man-in-the-middle (MITM) attacks.
To identify this type of behavior, we monitor sessions that (a) included untrusted certificates, and (b) were not terminated immediately after the handshake with the RST flag. 
Such sessions strongly indicate that the user has accepted the untrusted certificate (B5 in Appendix~\ref{sec:taxonomy_table}).

\subsubsection{Deep packet inspection}
Transmitting private information via an unencrypted protocol can expose the user to eavesdropping.
Thus, the existence of private content within the unencrypted network traffic of a user can be extremely indicative of the user's ISA. 
Deep packet inspection is a type of data processing which provides in depth inspection of the content of network traffic in order to extract meaningful patterns or insights.
This includes protocol headers, data structures, as well as the payload of messages (i.e., layers 2 to 7 of the OSI model).

In order to determine the user's likelihood of being the target of eavesdropping, we implement a dedicated DPI process. 
Specifically, we reconstruct the TCP session, decode GZIP/XML/JSON encoded traffic, and extract the content in a readable format. 
This content is then analyzed to identify personal information such as email addresses, passwords, credit card numbers, GPS coordinates, etc. (B3 in Appendix~\ref{sec:taxonomy_table}).
We also identify file downloads and use third party services (such as VirusTotal~\cite{virustotal}) to detect malicious files (B2 in Appendix~\ref{sec:taxonomy_table}).
Furthermore, focusing on the downloads of Android application packages (APK), we identify installations of applications from dubious sources (AI1 in Appendix~\ref{sec:taxonomy_table}).

\subsubsection{\label{subsubsec:contextual-analysis} Contextual analysis}
The security awareness of subjects is also affected by the context of use. 
Therefore, some ISA criteria reflect the behavior of users is specific contexts.
For instance, inserting private information into pop-ups (B4 in Appendix~\ref{sec:taxonomy_table}) and clicking malicious links while reading emails (VC2 in Appendix~\ref{sec:taxonomy_table}).
These criteria, which are indicative of ISA, cannot be evaluated directly from the raw network traffic due to encryption. 
However, in some cases we can infer that such behavior takes place using indirect indicators. 
For instance, in order to directly infer that a user clicks on malicious links while reading emails, it is necessary to analyze the subject's communication with email services which are typically encrypted.
Nonetheless, instances in which communication with email services occurred around the same time as communication with a malicious domain can be used as an indicator that the subject tends to behave carelessly while reading emails.
The values of the criteria measured through this method correspond to the number of appearances of the relevant attributes.

\subsection{\label{subsec:security-questionnaire} Security Questionnaire}
Security questionnaires are the most common method for assessing ISA in the literature.
The proposed security questionnaire was specifically designed to measure the self-reported behavior of subjects with respect to the security awareness criteria presented in~\cite{bitton2018taxonomy}.
When designing the survey, we followed the methodologies and questions presented in previous studies~\cite{androulidakis2011survey, mylonas2013delegate, esmaeili2014assessment, egelman2015scaling}, which pointed out two important aspects of behavior with respect to information security: \emph{preventive behavior} i.e., actions that the user performs to reduce the risk of being exposed to an attack; and \emph{confronting behavior} i.e., actions that the user performs when faced with a security risk.

Similar to previous studies, we formulated three type of questions:
\begin{enumerate}
\item Questions which focus on the \textit{likelihood} of taking an action in the context of specific ISA criteria.

\item Questions which focus on the \textit{frequency} at which the subject performs an action in the context of specific ISA criteria.

\item Questions which focus on \textit{management} of the device’s built-in connectivity.
\end{enumerate}

The complete questionnaire, together with the distribution of each answer are presented in Appendix~\ref{sec:behavior_qesutionnaire}.
In order to evaluate the questionnaire we conducted a pilot study, in which 20 users were requested to answer the questionnaire as well as participating in focus groups.
The main insight from this pilot-study was that illustrations and snapshots need to be added in order to make the questions more clear (e.g., security alerts, popups, permissions, encrypted websites and etc.)

In order to evaluate the criteria using the security questionnaire, we created an ordinal five-level scale for each question, where the first level represents a very reckless behavior, and the fifth level represents the most cautious behavior. 
This scale was considered the raw value of a user for a criterion (because some criteria are evaluated using multiple questions, the values of these criteria were calculated by averaging the raw values of the relevant questions). 
Then, we calculated the normalized value of a user for a criterion by normalizing the raw values of all users for the same criteria. 
The normalized value was considered as the value of criteria for the specific user.

\subsection{Computing the User's ISA Score for an Attack Class}\label{subsec: attack_models} 
The process of computing the ISA score of a smartphone user for an attack class include two inputs (see Figure~\ref{fig:score-calc}):
(1) a security awareness model for the particular class of attacks as defined by Bitton \etal~\cite{bitton2018taxonomy}; and (2) the values of criteria for a particular user as measured by the various data sources.
The three security awareness models (referred to as application model, MITM model, and phishing model) are presented in Appendix~\ref{sec:attack_models}, Figure \ref{fig:attack_models}, where we present the weight (y-axis) for each criteria (x-axis). In addition, for each model we map the three data sources to the criteria they measure. 
As can be seen, some of the data sources are able to measure only a partial set of criteria.
We attribute this to the fact that measuring the criteria is not a trivial task, especially when using a passive data source (such as the network traffic monitor). 
For instance, in the application model we can see that the network traffic data source can measure criteria that describe $62\%$ of the model, where the mobile agent data source can measure criteria that describe $87\%$ of the model. 
On the other hand, in the phishing model the most important criteria can be measured using all the three data sources. 
It is worth mentioning that the questionnaire data source can measure all of the criteria, however as previously mentioned, this data source tends to be very subjective.

\begin{center}
\begin{figure}[t]
\centering
	\begin{subfigure}[[h!]{0.49\textwidth}
      \includegraphics[width=\textwidth]{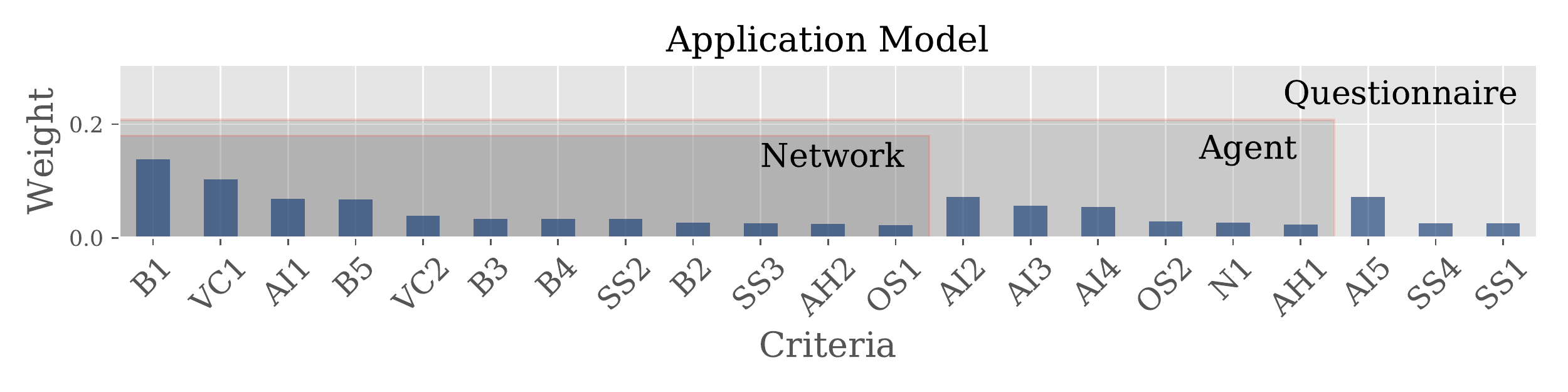}
      \label{fig:attack_model_Application}
      \vspace{-1.5\baselineskip}
    \end{subfigure}

    \begin{subfigure}[[h!]{0.49\textwidth}
      \includegraphics[width=\textwidth]{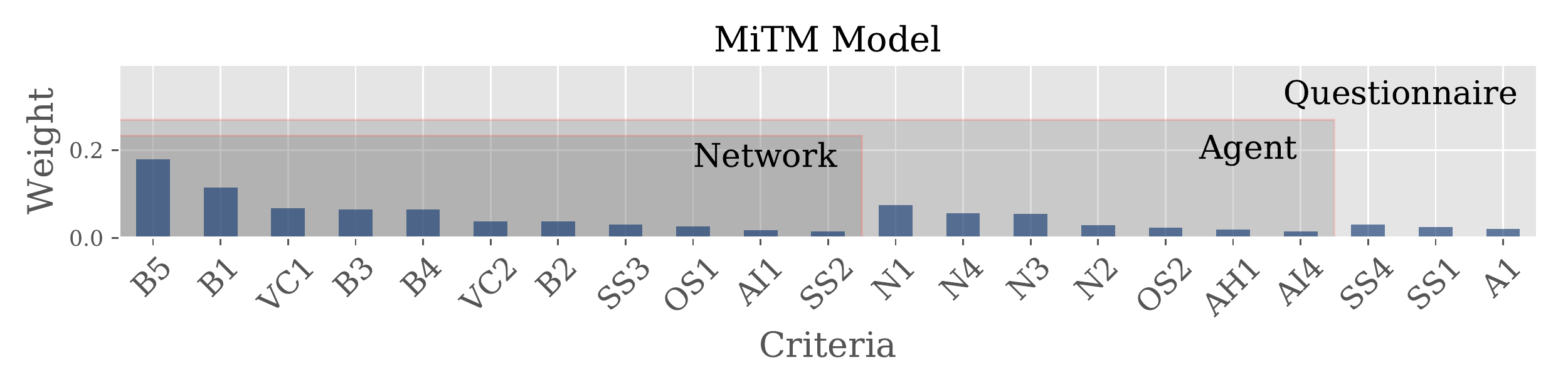}
      \label{fig:attack_model_MiTM}
      \vspace{-1.5\baselineskip}
    \end{subfigure}
	
    \begin{subfigure}[[h!]{0.49\textwidth}
      \includegraphics[width=\textwidth]{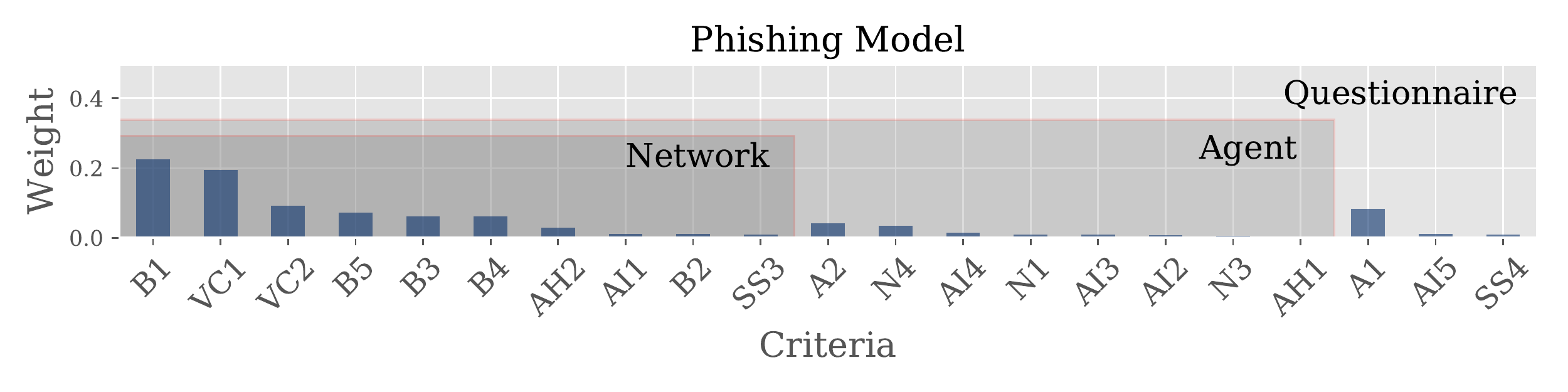}
      \label{fig:attack_model_Phishing}
      \vspace{-1.5\baselineskip}
    \end{subfigure}    
\caption{\label{fig:attack_models}The three attack models; the highlighted areas indicate the criteria that were measured using each data source.}
\end{figure}
\end{center}

The ISA score of a user $u$ for an attack class $a$, using data source $d$ is denoted by $s(u,a,d)$ and calculated as follows:

\begin{equation}
\label{eq:isa}
	s(u,a,d) = \frac{\sum_{i \in C_d}w_i(a) \cdot c_i(u)}{\sum_{i \in C_d}w_i(a) }
\end{equation}

where $C_d$ denotes the set of ISA criteria measured using the data source $d$, $w_i(a)$ denotes the weight of criterion $i$ in the awareness model $a$ and $c_i(u)$ represents the value of criterion $i$ as measured using data source $d$.

Given the ISA score, we can partition the  users into groups according to three security awareness levels denoted as low, medium, and high as follows: 

\begin{equation}
\label{eq:groups}
		L(u,a,d) = 
		\begin{cases}
		low, &s(u,a,d) < \mu-\beta \cdot \sigma \\ 
		medium, &\mu-\beta \cdot \sigma  \leq s(u,a,d) \leq  \mu+\beta \cdot \sigma \\
		high, &\mu+\beta \cdot \sigma  < s(u,a,d)
\end{cases}
\end{equation}

where $\mu$ and $\sigma$ denote the mean and the standard deviation of the ISA scores in the population and $\beta$ is a hyperparameter which can be used to adjust the sizes of the groups within the partition.

%% file: Resources/Sensors_list.tex
\begin{table*}[t]
\scriptsize
\newcolumntype{M}[1]{>{\centering\arraybackslash}m{#1}}
\begin{footnotesize}
\begin{tabular}{ |M{0.07\textwidth} | M{0.13\textwidth} | M{0.1475\textwidth} | >{\raggedright\arraybackslash}p{0.5525\textwidth}|}
\hline
\textbf{Group} & \textbf{Sensor} & \textbf{Sampling Method} & \textbf{Description}\\ 
\hline
\multirow{3}{*}{Connectivity} 
 & Wi-Fi  & When Wi-Fi  changed & Records the connected Wi-Fi access point (SSID, BSSID) and its security capabilities\\ \cline{2-4}
 & Bluetooth & When Bluetooth changed & Detects connected Bluetooth devices \\ \cline{2-4}
 & Traffic & Every 30 minutes & Collects statistic information about the volume of ingoing/outgoing network traffic of each package and process in the mobile device\\ \hline
\multirow{3}{*}{ Application} 
 & Installed apps & Every month & (1) Samples the installed applications and their permissions, and (2) scans the installed packages in VirusTotal and issues a notification when a package is updated or removed\\ \cline{2-4}
 & Running apps & Every 30 minutes & Samples the running applications and processes\\ \cline{2-4}
 & Application changes & When application changes & Monitors application updates, application deletions, and application installations.\\
\hline
\multirow{3}{*}{Content} 
 & Browser search & Every 30 minutes & Monitors the browser searches and URL's, scans the URL's in VirusTotal and Web Of Trust\\ \cline{2-4}
 & Emails & Every four hours & Provides statistics about the mailboxes (email sent, email received, spam received, etc.)\\ \cline{2-4}
  & SMS & Upon receiving SMS & Monitors links sent in SMSs. \\
\hline
\multirow{4}{*}{Device} 
 & Hardware & Every week & Collects model and brand information\\ \cline{2-4}
 & Software & Every week & Collects information about the OS, build number, and firmware\\ \cline{2-4}
 & Root checker & Every week & Checks whether the device is rooted\\ \cline{2-4}
 & Screen lock & Every week & Checks the screen lock type (PIN, pattern, none, etc.)\\ 
 \hline

\end{tabular}
\end{footnotesize}
\caption{\label{tab:agent-sensors-list}The sensors implemented within the mobile agent, classified to four groups.}
\end{table*}

%% file: evaluation.tex
\section{\label{sec:evaluation}Evaluation}

The proposed framework computes the ISA score of a user for a given SE attack class. 
The main hypothesis of this research is that this score will indicate the ability of the user to mitigate SE attacks. 
In other words, we assume that subjects with a high ISA score will be less likely to fail in mitigating SE attacks, while subjects with a low ISA score will be much more vulnerable to these attacks.
In order to evaluate the proposed framework, we conducted a long-term experiment involving 162 students who use their smartphones regularly.
During the experiment, we monitored the network traffic of the subjects, measured their behavior while operating their smartphones, and asked them to answer the security questionnaire. 
Towards the end of the experiment we exposed the subjects to four SE attacks. 
Then, we used the framework to calculate the ISA score of each subject, and evaluated the correctness of the score by cross-checking the score with the subject's performance in the challenges. 
In this section, we provide a detailed explanation of the evaluation process and results.

\subsection{\label{subsec:attacks-simulations}Social Engineering Challenges}
In the interest of providing a practical evaluation of the proposed framework, we implemented security challenges that resemble different SE attacks.
The primary advantage of this kind of evaluation is that it measures the ability of a subject to handle a real-life attack scenario.
The challenges are: phishing, spam pop-ups, suspicious permission requests, and certificate manipulation.
The last three challenges were implemented as part of the mobile agent, either by presenting a message or by presenting a pop-up Web-page. 
Only the first challenge (spear phishing) was initiated via an email.
In order to validate that a participant completed the first challenge via the mobile phone, we checked on our phishing web page the user-agent field (in the HTTP request), which indicates whether the client is a mobile-client or pc-client.
The users that complete the challenge using their PC were filtered out in the analysis of this challenge.

\subsubsection{\label{subsubsuc:phishing_attack}Spear phishing}
One of the most common SE attacks is phishing. 
The spear phishing challenge (presented in Figure~\ref{fig:phishing})  is a Web page that mimics a login page for student services.\footnote{This challenge was executed with the permission of the university's IT department and implemented under the department’s supervision.}
The attack scenario began by sending an email to the subjects that includes a message in which the subjects were asked to authenticate themselves using an attached link due to a problem with the university's computer systems. 
The sender of the email was ``STUDENT ADMINISTRATION," a sender that the subjects are familiar with as emails associated with the university's administration are sent by this entity.
The email was sent to the participants during the academic semester, where administrative emails are more likely to be sent by the university's IT department.
Although the email looked like a legitimate email, there were several indicators that it was a phishing scenario. 
First, the phishing email was not sent from the university's mail system. 
Second, the attached URL was not associated with the university's domain. 
Third, the phishing Web page was not connected via the HTTPS protocol.
To preserve the subjects' privacy, the authentication information was not transmitted to the server. 
Instead, we implemented a client side script that validates the username and ID number.
Please refer to Section~\ref{subsec:ethics} for further issues related to privacy and ethical considerations. 
Within this challenge, a subject who click on the malicious URL, and insert his/her private credentials in the phishing Web site was considered as failed in mitigating the attack.

\subsubsection{\label{subsubsuc:spam_attack}SPAM pop-ups}
Spam can expose the subject to privacy and security threats. 
The spam challenge (presented in Figure~\ref{fig:spam}) is a Web page which offers a chance to win an iPad by answering a simple question; the subject is requested to expose personal information (such as their first name, date of birth, and email address).
This attack scenario is executed when the user is within the context of mobile browsing, with a pop-up that was triggered by the mobile agent and appears on the subject's screen.
Similar to the phishing challenge, the personal information supplied by the subjects was not transmitted to the server.
Within this challenge, a subject who insert his/her private information to the spam pop-up was considered as failed in mitigating the attack.

\subsubsection{\label{subsubsuc:permission_attack}Suspicious permission request}
Malicious applications can trick unaware users into granting dangerous permissions during runtime.
In this challenge (presented in Figure~\ref{fig:super-user}), the Web page contained a request to grant superuser privileges for a "browser plug-in." 
The attack scenario was initiated when the user is within the context of mobile browsing; this triggered a "system alert" which was initiated by the mobile agent and appeared on the subject's screen. 
The subject could reject the request approve it forever, or approve it only for that time. 
Within this challenge, a subject who grant the superuser privileges to the browser plug-in was considered as failed in mitigating the attack.

\subsubsection{\label{subsubsuc:certificate_attack}Certificate manipulation}
\begin{center}
\begin{figure*}[t]
    \centering
    \begin{subfigure}[t]{0.24\textwidth}
        \includegraphics[width=\textwidth,keepaspectratio]{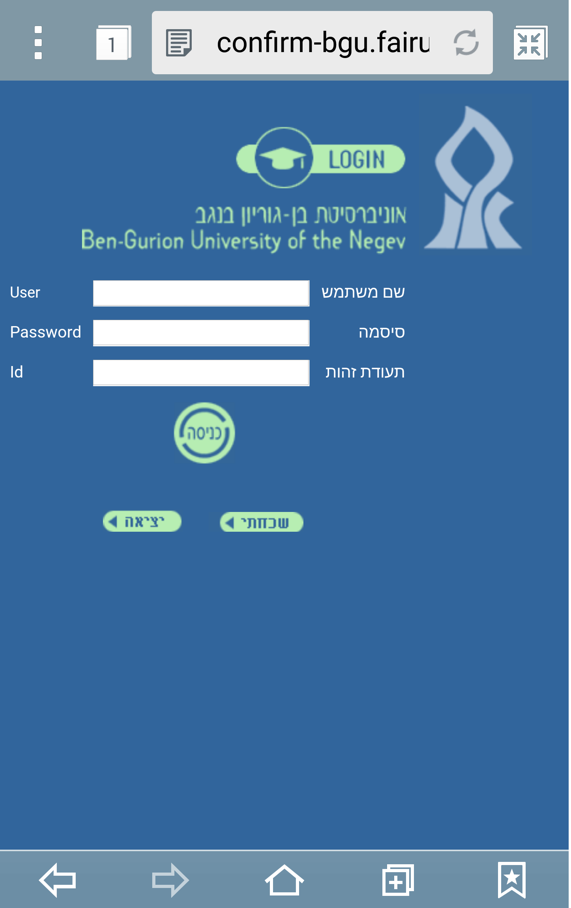}
        \captionsetup{justification=RaggedRight}
        \caption{The spear phishing challenge Web page.}
        \label{fig:phishing}
    \end{subfigure}
    \begin{subfigure}[t]{0.24\textwidth}
        \includegraphics[width=\textwidth,keepaspectratio]{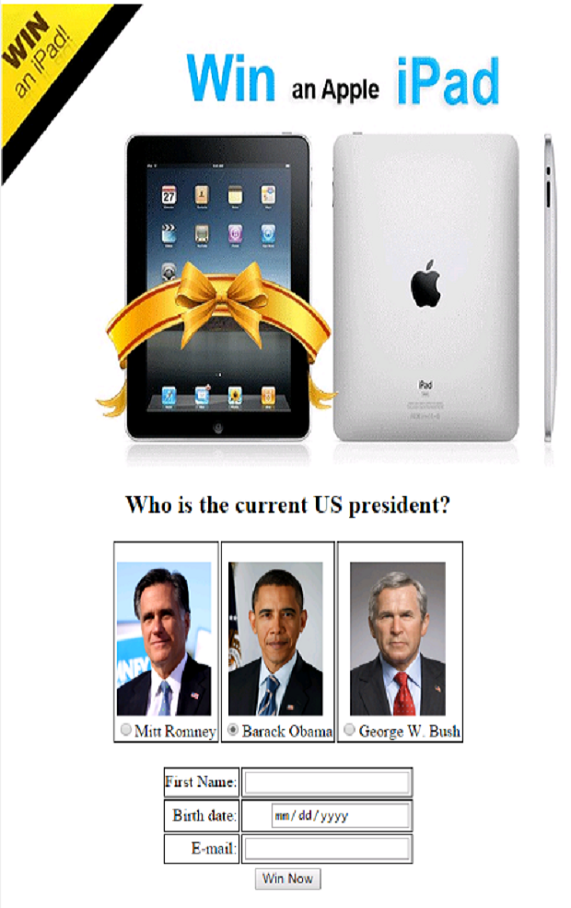}
        \captionsetup{justification=RaggedRight}
        \caption{The spam challenge offering a chance to win an iPad.}
        \label{fig:spam}
    \end{subfigure}
        \begin{subfigure}[t]{0.24\textwidth}
        \includegraphics[width=\textwidth,keepaspectratio]{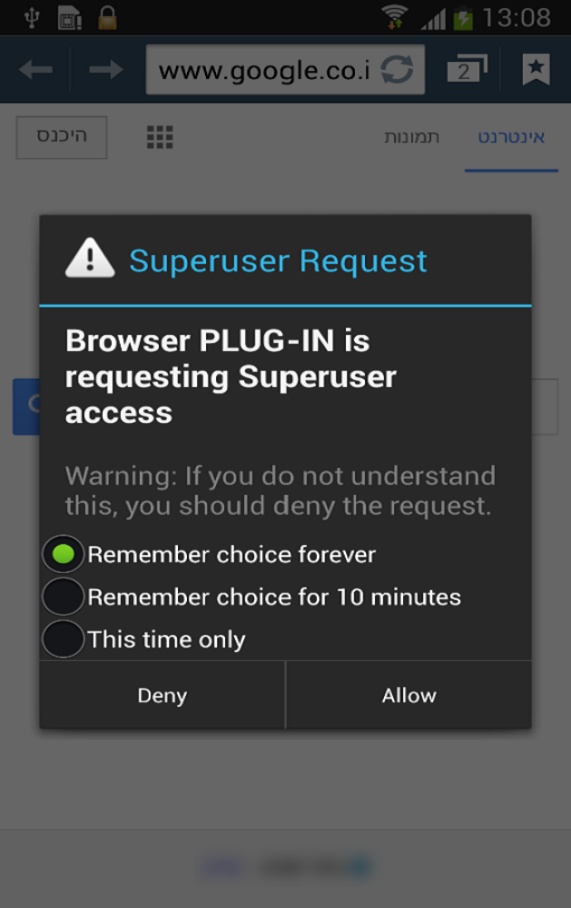}
        \captionsetup{justification=RaggedRight}
        \caption{Request to grant super user privilege for a browser plug-in.}
        \label{fig:super-user}
    \end{subfigure}
    \begin{subfigure}[t]{0.24\textwidth}
        \includegraphics[width=\textwidth,keepaspectratio]{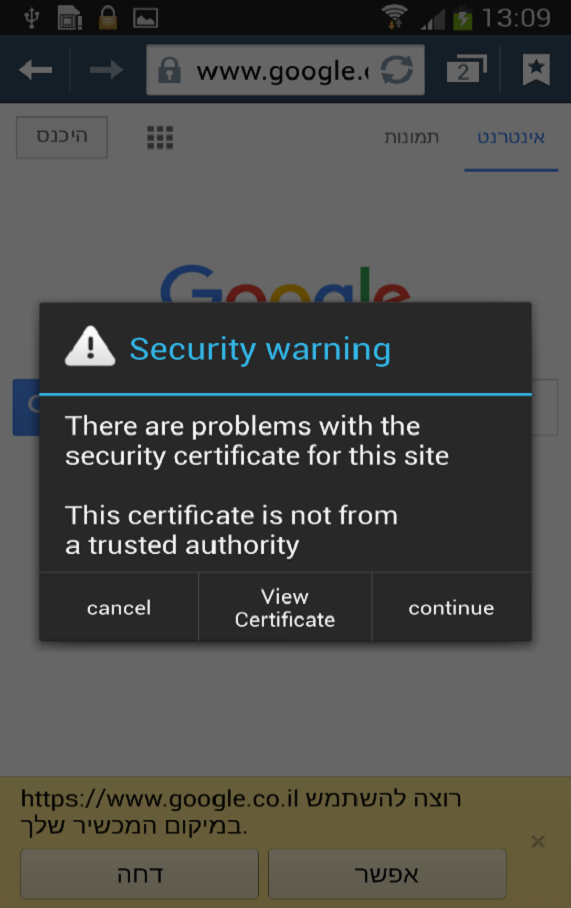}
        \captionsetup{justification=RaggedRight}
        \caption{Certificate manipulation challenge.}
        \label{fig:certificate}
    \end{subfigure}
    \caption{The social engineering challenges} \label{fig:challenges1}
\end{figure*}
\end{center}

In this challenge (presented in Figure~\ref{fig:certificate}), the Web page contained a security warning about using a certificate that is signed by an untrusted certificate authority. 
The attack scenario was initiated when the subject opened a secure connection to a website by using the browser.
Then, triggered by the mobile agent, a "system alert" was issued and appeared on the subject's screen.
The subjects could accept the certificate, request more information, or cancel and navigate out of the suspicious website; the "more information" button provided information written in Chinese. 
Within this challenge, a subject who accept the untrusted certificate was considered as failed in mitigating the attack.

\subsection{\label{subsec:data-collection}Data Collection}
\subsubsection{Recruitment}
\review{
Collecting data from real subjects in their natural environment and for a long-term period is very challenging.
In order to recruit a diverse population of subjects, we sent emails to \emph{active students} and \emph{graduate students} of two different academic institutes in Israel.
The title of the email was "Invitation to participate in a study on Android smartphones."
Within the content of the email the potential subjects were told that the purpose of the study is to explore their browsing habits (we intentionally obfuscated the fact that the study is about information security to avoid \emph{Hawthorne effects}, in which subjects modify their behavior in response to their awareness of being observed)\footnote{This form of deception was approved by the institutional Human Subjects Research Committee.}.
}

About 1000 subjects were agreed to participate in the study, we invited 300 of them to our laboratory in order to sign in to the experiment, receive guidance, and install the data collection framework.
About 250 of subjected received the guidance and 220 of them signed in to participate in the research. 
Upon signing the consent form, subject were requested to complete a demographic survey on Google Forms.
Each one of them received a compensation of 200 USD.

It should be mentioned that during the first week of the experiment 40 participant opted to withdraw from the experiment; this was due to a high battery consumption or incompatibility of the data collection framework (which resulted in repeating crashes of the agent).
%The main reasons for that drop out were igh battery consumption and incompatible device, which cause the data collection framework to crash.
In addition, we filtered out 18 participants who did not activated the data collection framework regularly.
Finally, 162 participants who use their smartphones regularly served as the subjects of the experiment and for the data analysis and evaluation.

\begin{figure}[t]
\centering
    \begin{subfigure}{\linewidth}
        \centering
        \includegraphics[width=1\textwidth]{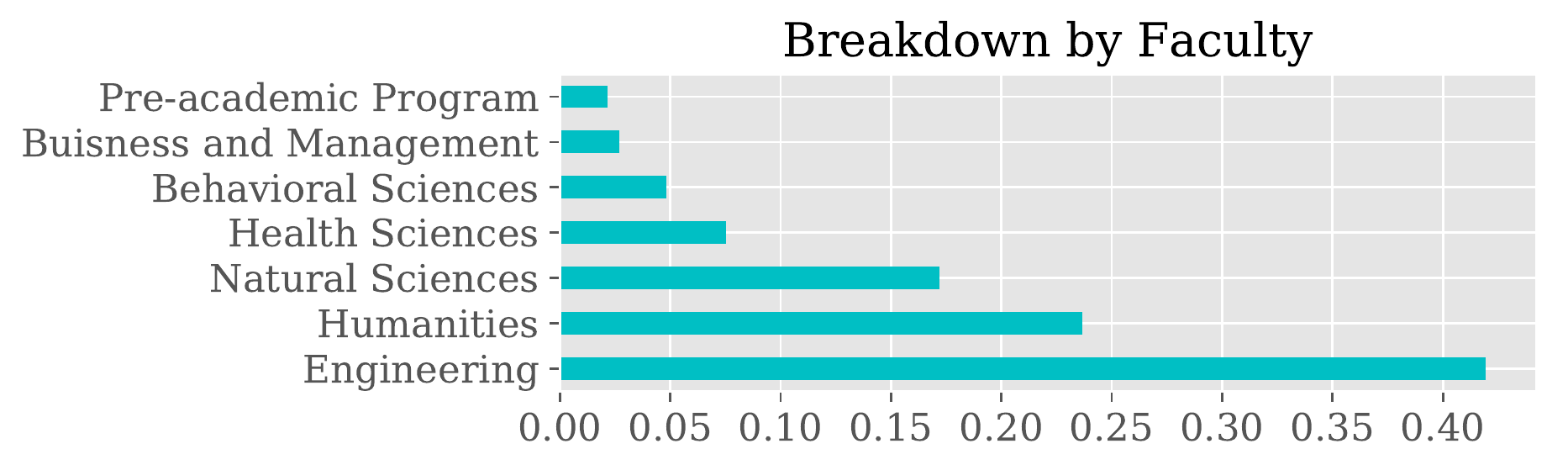}
    \end{subfigure}
\begin{subfigure}{0.45\linewidth}
        \centering
        \includegraphics[width=\textwidth]{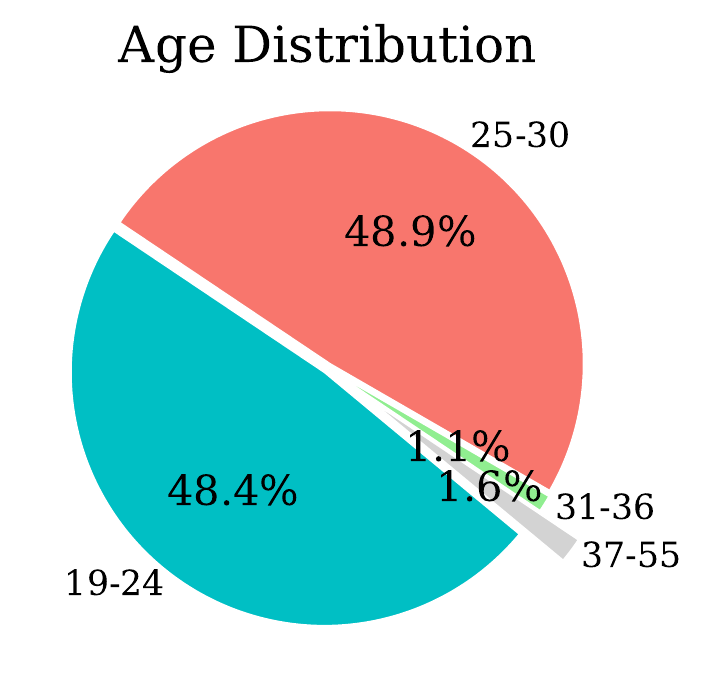}
    \end{subfigure}
\begin{subfigure}{0.45\linewidth}
        \centering
        \includegraphics[width=0.98\textwidth]{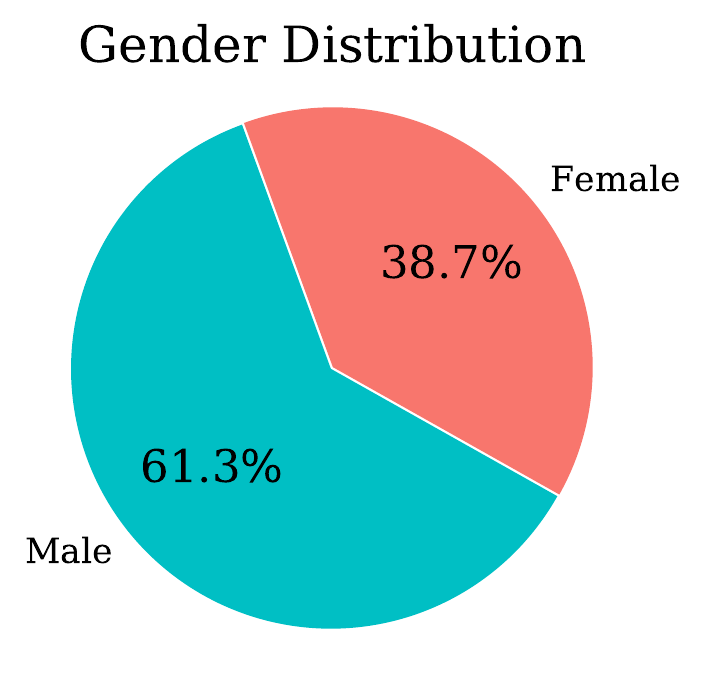}
    \end{subfigure}

    \caption{\label{fig:demographic_distribution}Demographic properties of the subjects.}
\end{figure}

\begin{figure}[t]
\centering
\includegraphics[width=0.98\linewidth]{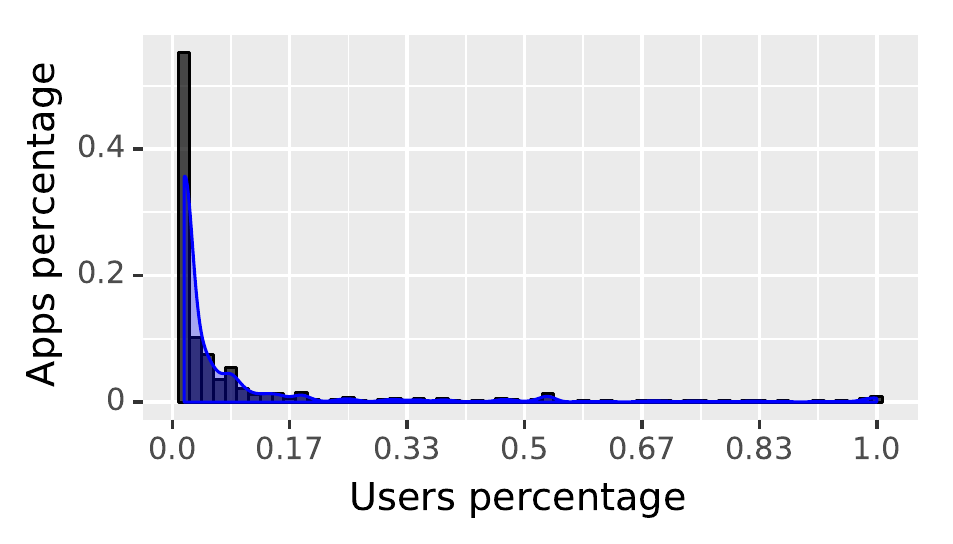}
\caption{Distribution of distinct installations.}
\label{fig:apps-distribution}
\end{figure}

Descriptive statistics of the population are presented in Figure~\ref{fig:demographic_distribution}.
As can be seen, the population consisted mainly of 19 to 30 year-old's, almost half of which are students/university-graduated from the faculty of engineering.
We also analyzed the distribution in the number of distinct application installations among the experiment population.
As can be seen in Figure~\ref{fig:apps-distribution}, most of the applications are installed on less than 20\% of the devices and more than 40\% of the applications are installed on one device only. 
On the other hand, popular applications such as Whatsapp, Chrome, Waze, Facebook, Shazam, Skype, GetTaxi, and Instagram were installed on more that 50\% of the devices (see Figure~\ref{sec:user-apps-install}).
This indicates that the population of participants in the experiment is diverse.

\begin{figure}[t]
\centering
\includegraphics[width=0.96\linewidth]{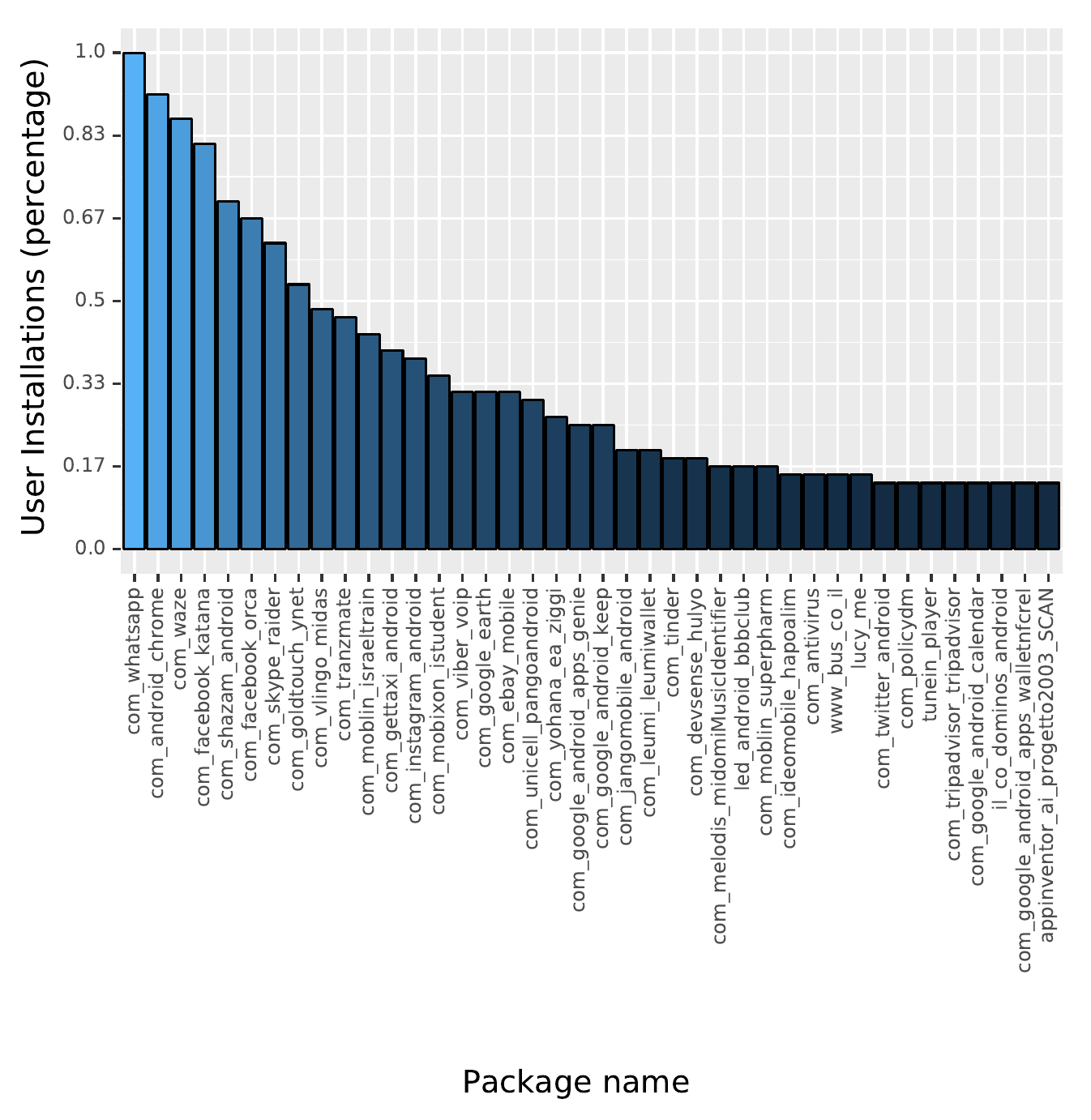}
\caption{Distinct installations of popular applications (after filtering system applications).}
\label{fig:user-apps-install}
\end{figure}

In order to validate that the users use their smartphone regularly throughout the study.
We routinely sent a "keep-alive" message to our data collection servers every 30 minutes. 
If for a period of 6 hours such message was not received the participant was automatically notified via an email.
We also routinely validated that the VPN connection is active. 
In the case that a user was not connected to the VPN for a period of six hours, he was automatically notified via an email.
Nevertheless, by the end of the experiment we requested each participant to complete the missing days by extending the experiment period respectively.
At the end, each participant was active for at least seven weeks.

\subsubsection{Experiment setup}
In the experiment, each subject was requested to install two Android applications: a dedicated Android application that was developed by the research group and used to passively monitor the subjects' behavior based on a set of sensors (as described in~Section~\ref{subsec:mobile-agent}); and a commercial VPN client application that was used to redirect the subjects' Internet traffic to the server used in the experiment. 
It should be mentioned that all of the data was transmitted securely to this server. 

The experiment was conducted according to the following schedule:

\begin{itemize}
\item \textbf{Opening session:}
At the beginning of the experiment the subjects were asked to install the applications on their personal smartphones and keep them there for a period of seven weeks.

\item \textbf{Passive data collection:}
During the first three weeks of the experiment, we passively monitored the subjects' behavior and Internet traffic using the mobile agent and VPN applications.
During this period the subjects were not exposed to any challenges.

\item \textbf{Active probing:}
From the fourth week on, each subject was exposed to a different challenge each week. 
To prevent the order of the challenges to influence the results, the order was based on the lattice square design~\cite{yates1940lattice}.

\item \textbf{Security questionnaire:}
At the end of the seventh week, the subjects were requested to answer an online security questionnaire.
\end{itemize}

\subsubsection{Privacy and ethical considerations}
\label{subsec:ethics}
The experiment involved collecting sensitive personal information from real subjects for a long period of time, including their browsing patterns.
We did our best to preserve the subjects' privacy and reduce any risks associated with participating in the experiment. 
The experiment was approved by the institutional review board (IRB), subject to the following precautions:

\begin{enumerate}
\item The subjects participated in the experiments, freely, at their own free will, and provided their formal consent to participate in the research.
The subjects received a one-time payment as compensation for their participation.
Although, at the beginning of the experiment the subject were not aware to its purpose, they were fully aware of the type of data that would be collected and were allowed to withdraw from the study at any time.

\item Anonymization was applied to the data. 
At the beginning of the experiment, a random user ID was assigned to each subject, and this user ID served as the identifier of the subject, rather than his/her actual identifying information. 
The mapping between the experiment's user ID and the identity of the subjects was stored in a hard copy document that was kept in a safe box; at the end of the experiment we destroyed this document.

\item During the experiment, the communication between the agents and the servers was fully encrypted and the collected data was stored in an encrypted database.
At the end of the experiment the data was transferred to a local server (i.e., within the institutional network), which is not connected to the Internet. 
Only anonymized information of the subjects was kept for further analysis.
\end{enumerate}

\subsection{\label{subsec:evaluation-method}Statistical Analysis}
The main hypothesis of this research is that subjects with high ISA score are likely to mitigate SE attacks, while subjects with low ISA score will fail doing so.
We evaluate this hypothesis with respect to the three data sources
($d \in \{questionnaire, network, agent\}$) 
using the four challenges presented in Section~\ref{subsec:attacks-simulations} ($a\in \{phishing, spam, permissions, certificate\}$)
In order to evaluate the above hypothesis we start with the $\chi^2$-test to check whether the failure ($f(u,a)$) of the subject $u$ in challenge $a$ and his/her awareness level ($L(u,a,d)$) are statistically dependent.  
In a standard $\chi^2$-test the null hypothesis states that the two variables are independent.
If the null hypothesis is accepted then the ISA score computed based on the specific data source does not reflect the subject's ability to mitigate the challenge. 
If the alternative hypothesis is accepted, that is, success in the SE challenges and the level of security awareness are statistically dependent, then the latter can be used to guide security awareness policies. 

However, $\chi^2$-test does not provide the complete picture. 
For example, if most people with high security awareness will fail mitigating the SE attacks and most people with low security awareness will succeed doing so, then both variables will be statistically dependent but the computed ISA score will be misleading. 
We will demonstrate such situation in Section~\ref{subsec:practical-results} based on the data collected during our experiment. 

In order to complement the statistical analysis we use Pearson correlation coefficient to check for trends and derive actionable insights. 
We define the \emph{average ISA score} ($\overline{s}(l,a,d)$) of the subjects with security awareness level $l\in\{low, medium, high\}$ as: 
$$\overline{s}(l,a,d) = \text{avg}_{u: L(u,a,d)=l}\{s(u,a,d)\}.$$ 
We define the \emph{success rate} ($sr(l,a,d)$) of the subjects with security awareness level $l$ as the fraction of the subjects that succeeded at mitigating the challenge out of all subjects with security awareness level $l$: 
$$sr(l,a,d)=\frac{\left|\{u : L(u,a,d)=l \wedge f(u,a)=success\} \right|}{\left|\{u : L(u,a,d)=l \} \right|}$$

We examine the Pearson correlation coefficient between the two random variables $sr(l,a,d)$ and $\overline{s}(l,a,d)$ across different security levels.  
A high positive correlation along with significant statistical dependency according to $\chi^2$-test will suggest that the level of ISA computed based on the source $d$ correctly reflect the ability of the subjects to mitigate SE attacks similar to $a$.    
Ideally, the following inequality should hold for any $a$ and $d$: $sr(low,a,d)<sr(medium,a,d)<sr(high,a,d)$.

\subsection{\label{subsec:practical-results}Results}
\begin{figure}[t]
%\vspace{2mm}
	\centering
  \begin{subfigure}[t]{.98\linewidth}
    \includegraphics[width=\linewidth]{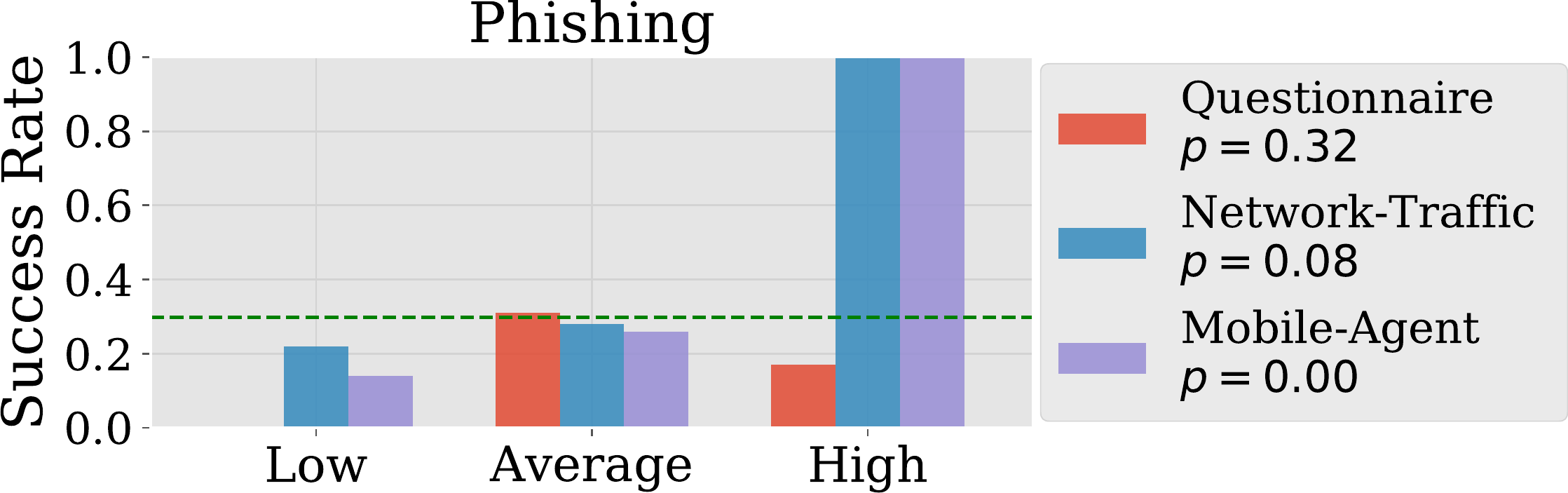}%[1ex]%
  \end{subfigure}
  \vspace{1mm}
  \begin{subfigure}[t]{.98\linewidth}
    \includegraphics[width=\linewidth]{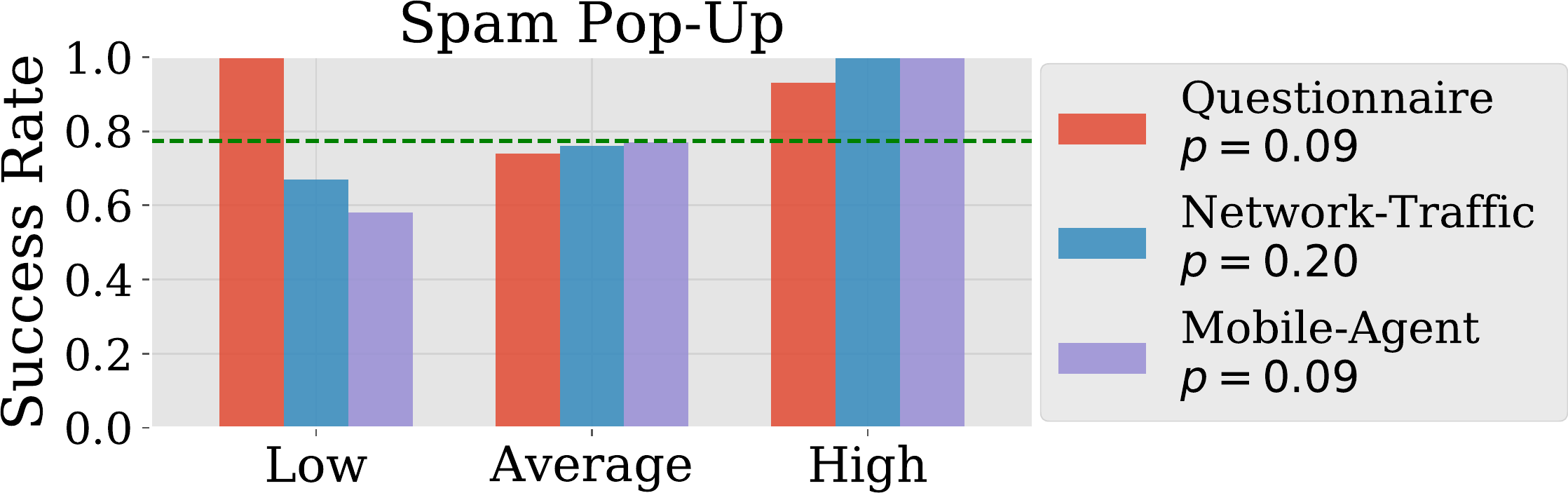}%[1ex]%
  \end{subfigure}
  \vspace{1mm}
    \begin{subfigure}[t]{.98\linewidth}
    \includegraphics[width=\linewidth]{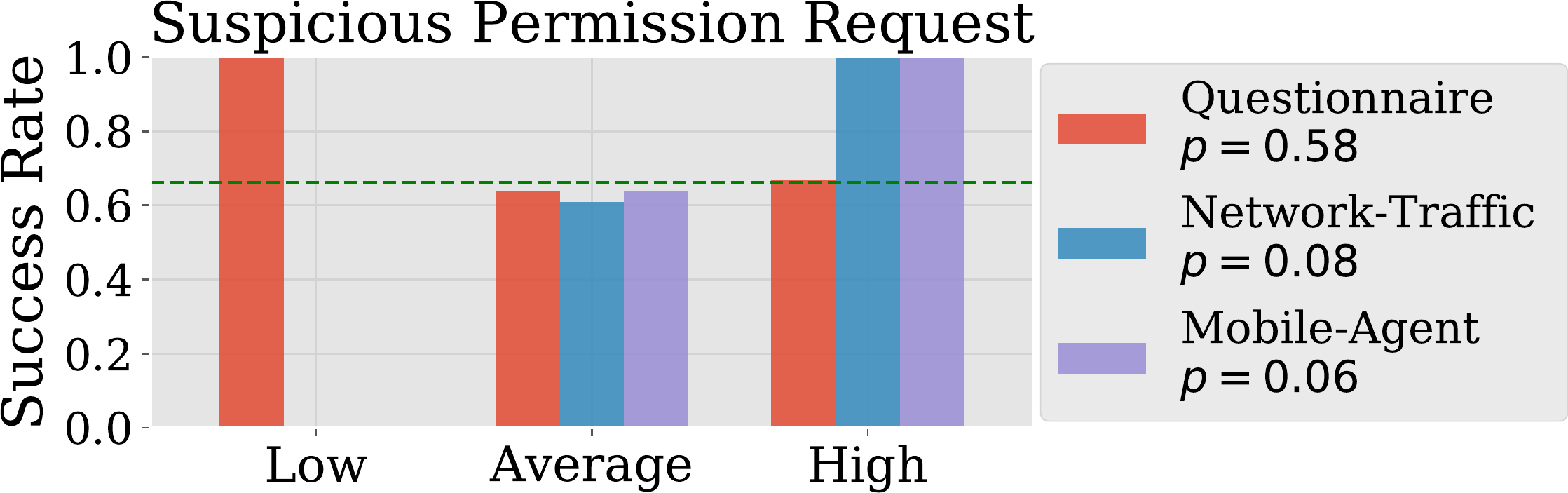}%[1ex]%
  \end{subfigure}

    \begin{subfigure}[t]{.98\linewidth}
    \includegraphics[width=\linewidth]{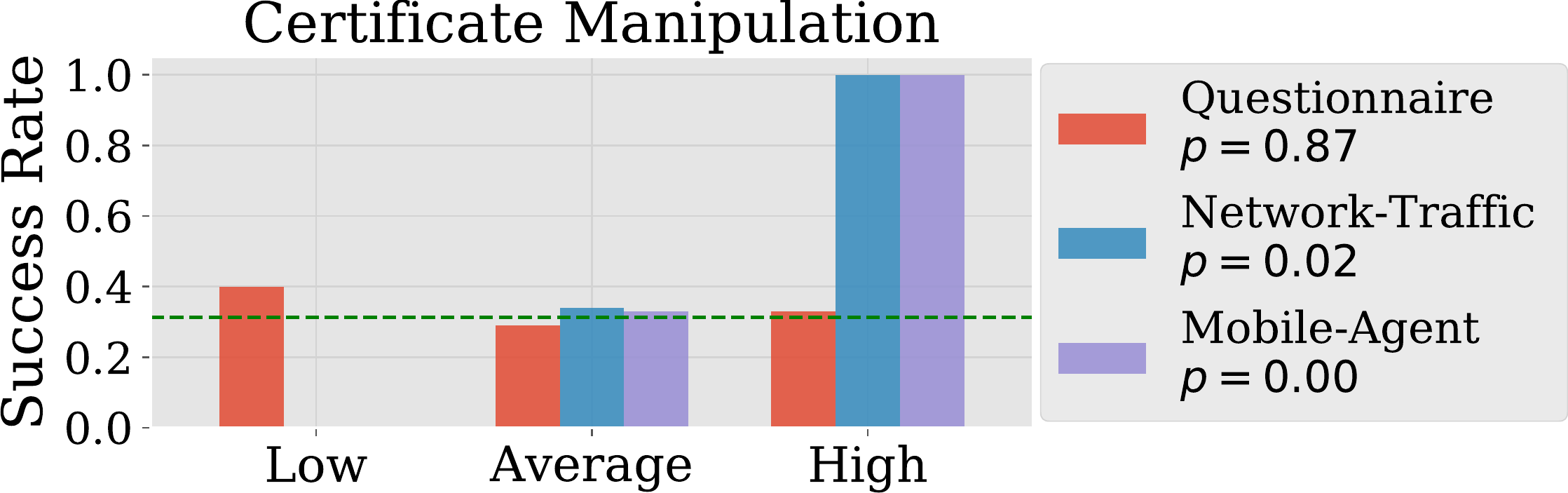}%[1ex]%
  \end{subfigure}

\caption{\label{fig:success_rates} The success rate of each awareness group (low, average, high) in each challenge; the dashed line represents the overall success rate for the challenge.}  
\end{figure}

The statistical analysis described above was applied on the three data sources (questionnaire, network and agent).
Success rates of subjects are presented in Figure~\ref{fig:success_rates}, where each sub-figure present the performance of the three data sources for a specific challenge.
The $p$-values associated with the $\chi^2$ tests for each data source are presented in the figure legend.
The dashed lines indicate the overall success rate for each challenge.

\begin{figure}[t]
  \begin{subfigure}[b]{\linewidth}
  	\raggedleft
    \includegraphics[width=0.9\linewidth]{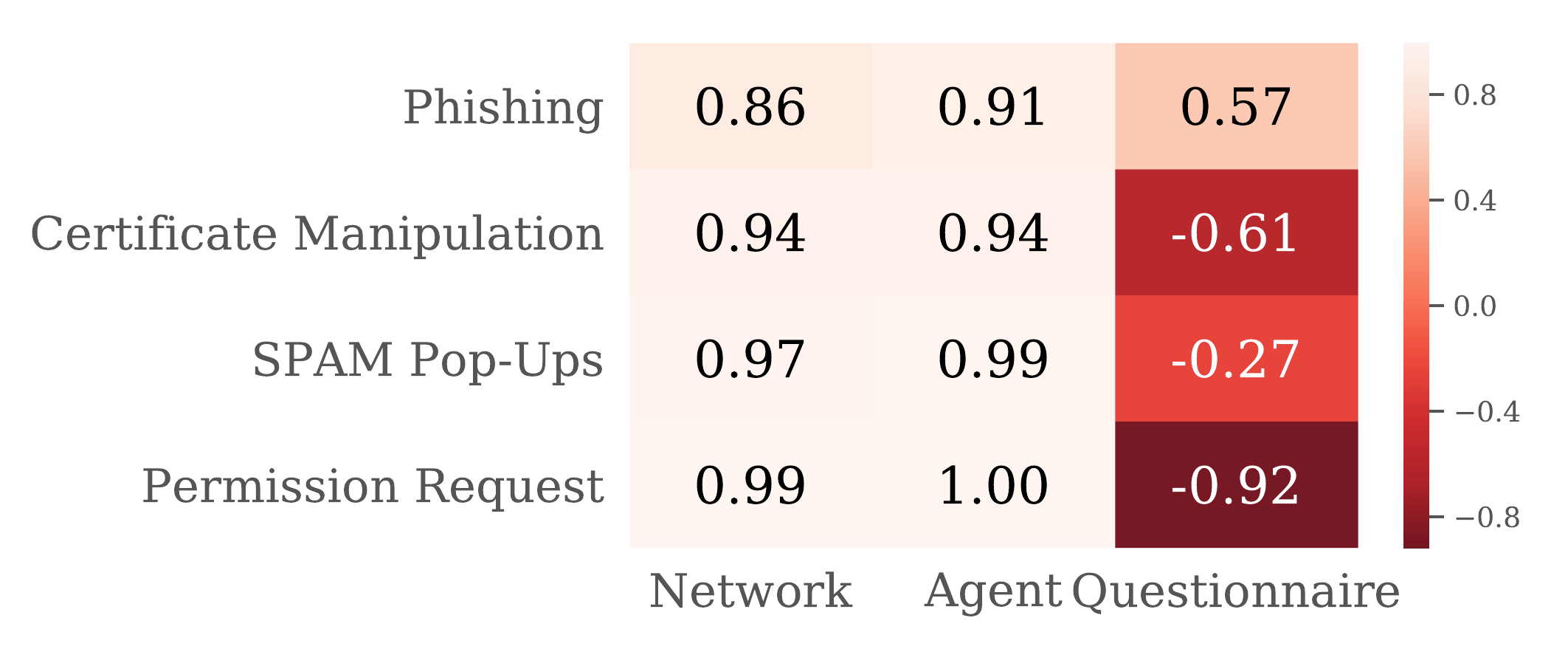}%[1ex]%
    \subcaption{\label{fig:correlation_scores_success_rate}\centering The correlation between the average awareness scores and the success rates for each challenge.}
  \end{subfigure}

  \begin{subfigure}[b]{\linewidth}
    \includegraphics[width=\linewidth]{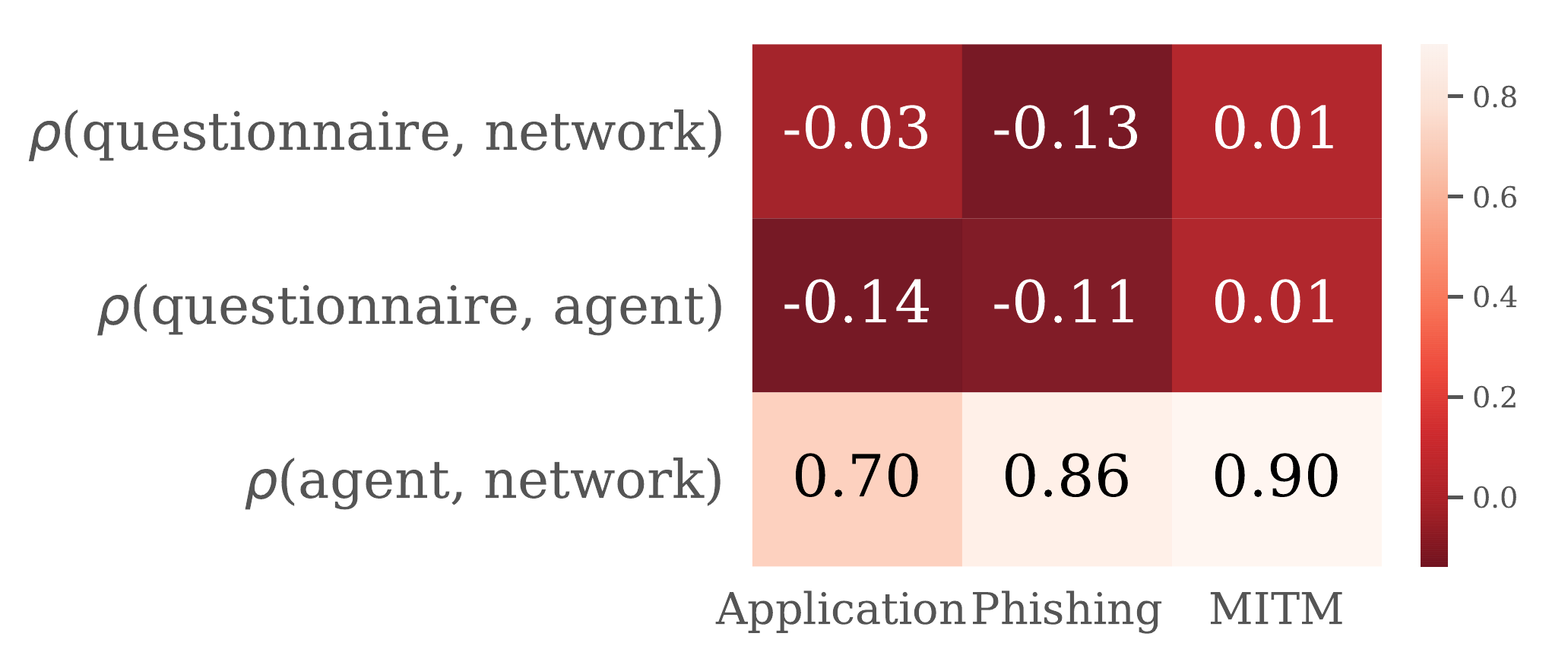}
    \subcaption{\label{fig:correlation_measuring_tools}\centering The correlation between the awareness scores generated from the different data sources for each awareness model}
  \end{subfigure}
\caption{\label{fig:correlation_tests} Results of the correlation tests.}
\end{figure}

As can be seen, in most of the challenges, the classification based on the mobile agent and network traffic analysis are statistically significant.
In addition, in all of the challenges, the mobile agent and network traffic monitor classify the subjects to awareness groups that are correlated to the success rates for the challenge.
This observation is also reflected in the correlation coefficient matrix presented in Figure~\ref{fig:correlation_scores_success_rate}, since the average ISA scores based on the mobile agent and network traffic analysis are linearly correlated with the success rates for the challenges.
On the other hand, the assessment of ISA based on the security questionnaire produced a negative correlation;
namely, the questionnaire classified the subjects into awareness groups that are not correlated to the success rates for the challenge.

This insight prompted us to further investigate the correlation between the ISA scores of different data sources.
In Figure~\ref{fig:correlation_measuring_tools}, we present the Pearson correlation coefficients for each pair of data sources (y-axis) and awareness model (x-axis). 
As can be seen, for the phishing and MITM models the ISA scores generated by the network traffic monitor and mobile agent are highly correlated.
We attribute this to the fact that both the phishing and MiTM models are based on criteria that can be measured precisely from the network and agent.
However, for the application model the correlation of the ISA scores generated by the network traffic and mobile agent is slightly lower.
We attribute this to the fact that some of the criteria are not measurable through the network.
On the other hand, the ISA scores generated by the security questionnaire are not correlated with the ISA scores derived from the network traffic monitor or agent data sources.

\begin{figure*}[t]
\centering
    \includegraphics[width=1\textwidth]{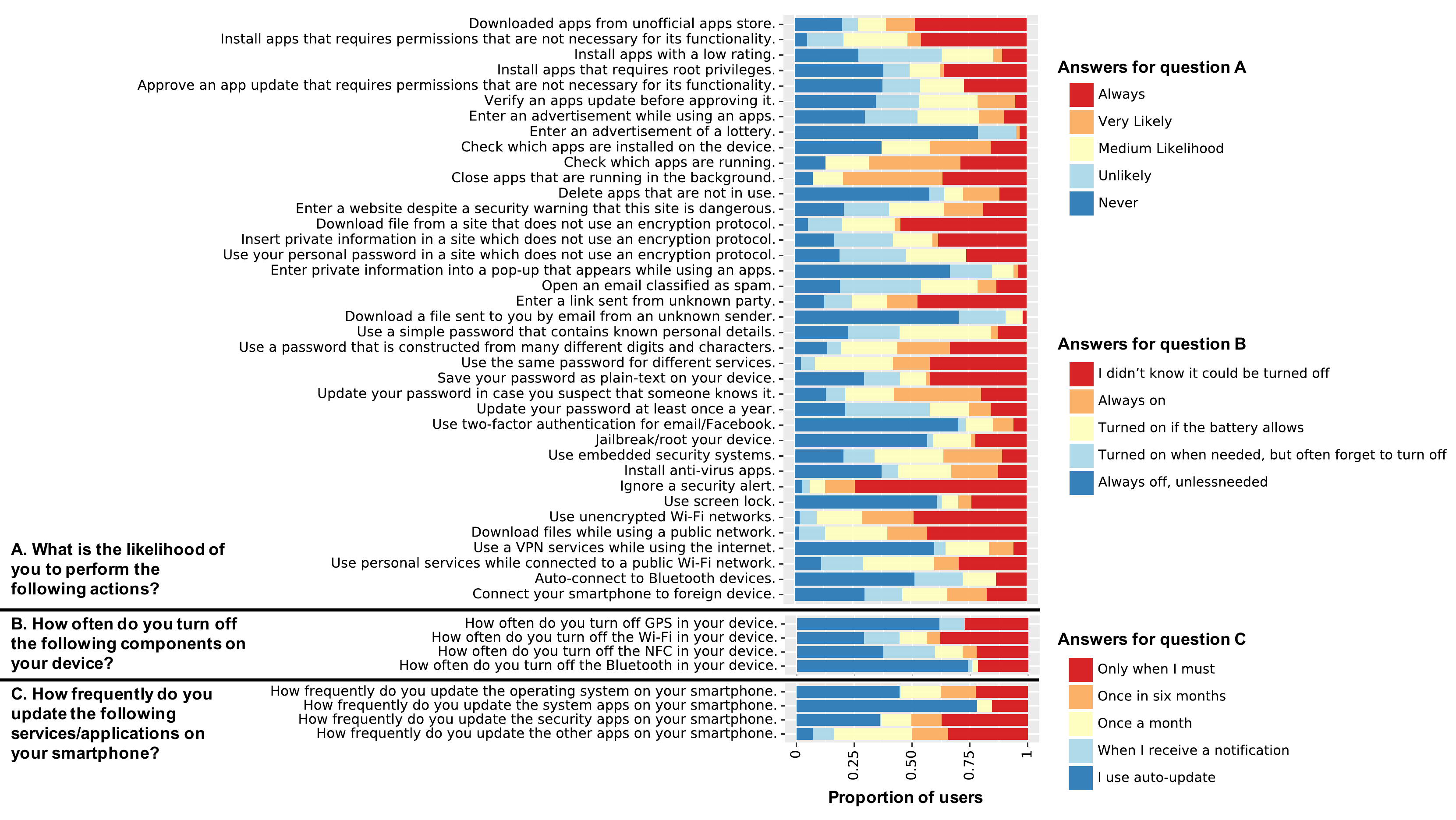}
    \caption{\label{fig:ans-distribution} Questionnaire answers' distribution}  
\end{figure*}

The contradictions between the self-reported behavior of the subjects (measured by the behavioral questionnaire) and the actual behavior (measured by the mobile agent, network traffic, and the challenges) can be seen more clearly in lower resolution by analyzing subjects' responses specific questions. \\
The distribution of questionnaire answers is presented in Figure~\ref{fig:ans-distribution}).

\begin{figure*}[t]
\centering
\includegraphics[width=1\linewidth]{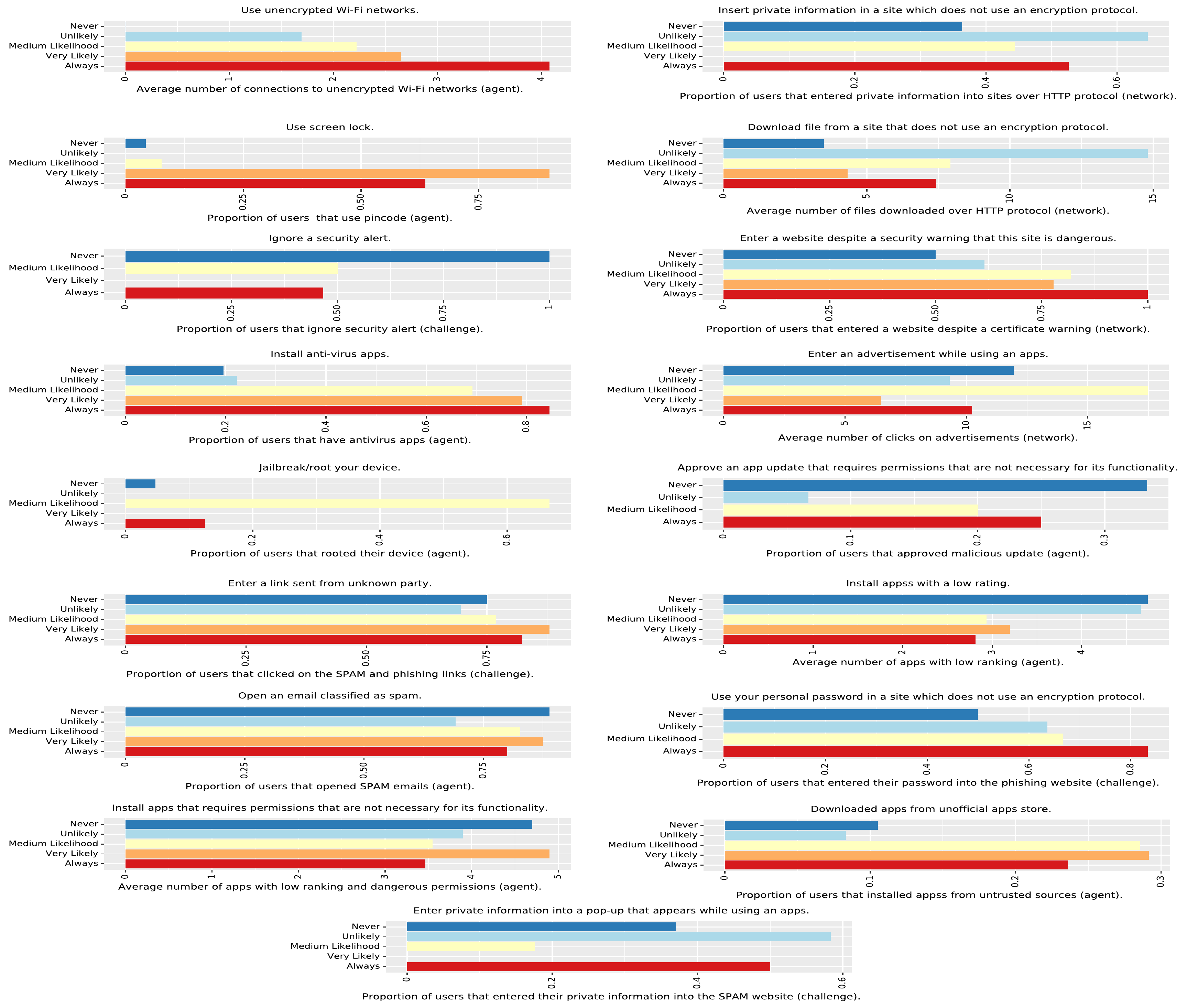}
\caption{\label{fig:cmp-ans-mesurments}Comparison of self-reported behaviour (questionnaire) to measurement data (agent, network-traffic and challenges).}
\end{figure*}

An interesting observation is that about $70\%$ of the subjects have reported that they would \textit{never} provide personal data in pop-up Web pages.
However, as presented in Figure~\ref{fig:cmp-ans-mesurments}, when analyzing the actual behaviour of these subjects we noticed that almost $40\%$ of them entered their private data in the SPAM website (our SPAM challenge).
Similarly, almost $60\%$ out of the subjects who reported that it is \textit{unlikely} that they would provide personal data in a pop-up Web pages did entered their heir private data in the SPAM website (our SPAM challenge).

Similar findings were observed while investigating questions related to certificates, lock screen protection, and file downloads.
For example, $50\%$ of the subjects who claimed that would \textit{never} enter a Website for which a security warning was displayed, failed in the certificate manipulation challenge; $35\%$ of the subjects who claimed that they \textit{always} use lock screen protection (e.g., password, PIN code, or pattern) did not use any method at all; and subjects who claimed that it is \textit{unlikely} that they would download a file via (unsecured/unencrypted) HTTP protocol, downloaded on average executable files via HTTP during the experiment.
On the other hand, subjects who reported that they always download files via HTTP, downloaded on average six executable files via HTTP during the experiment. 

In general, from the analysis of the gap between the self-reported behavior and the monitored behaviour (i.e., real measurements) we conclude that security aware subjects less determined in their questionnaire's answers, while subject that were found to be with a low security awareness level were more determined and reported a "safer" behavior.    
These conflicts supports the underlying assumption made in previous studies that the self-reported behavior is not reliable for measuring the behavior of subjects.

%% file: discussion.tex
\section{\label{sec:discussion}Discussion}
In this paper, we develop a framework for assessing the ISA of smartphone users.
The framework utilize three different data sources in order to evaluate an extensive list of criteria for a security aware smartphone user.
Although in previous sections we showed that the mobile agent data source produced the best results, there are additional criteria that should be considered when selecting a data source in production.
In this section we elaborate on those criteria.

\subsection{Interactivity} A very important aspect to consider when selecting a data source in production is the level of interaction with subjects. 
A high level of interactivity requires more resources, hence, it is less likely to be performed on an ongoing basis. 
In terms of this criterion, the questionnaire requires the most significant amount of interaction with each subject. 
In contrast, measuring the ISA via a mobile agent requires a significantly lower level of interaction with the subjects as it requires a one-time installation of the application on their devices. 
However, in this category, the network traffic monitor data source outperforms all other data sources, as it does not require any interaction with the subjects.

\subsection{Reliability} An unreliable data source will eventually produce an inaccurate assessment. 
Hence, reliability is an extremely important aspect to consider when selecting a data source. 
With regard to this criterion, the completion of a questionnaire, which is largely dependent on the subjects' reliability, has inherent disadvantages. 
As presented in the previous section, when analyzing the security questionnaire we observed that subjects behaved differently than their stated behavior, as reported in the security questionnaire. 
The high performance of an ISA assessment based on objective data sources compared to performance of the security questionnaire data source, indicates that in terms of reliability, objective measures such as the network traffic monitor and mobile agent are superior.

\subsection{Timeliness} Another important aspect to consider is the monitoring period required for a reliable assessment of the subject's ISA. 
Although, this aspect is less relevant for a security questionnaire, it is crucial for the other data sources, since a data source that is based on behavioral monitoring is more likely to be inaccurate without a sufficient amount of data. 
With respect to this criterion, the mobile device agent is superior to the network traffic monitor. 
The main reason for this is that many types of security criteria can be measured by the mobile device agent immediately, at installation time, by observing device parameters, whereas measuring the same criteria solely by utilizing network traffic may take more time. 
A good example of such criteria are those criteria that are based on the list of installed applications (e.g., the presence of an anti-virus application). 
Obtaining such information by using the mobile device agent is as simple as calling an API function, whereas assessing such information from the network traffic depends upon the presence of communication between the anti-virus application and its server.

\subsection{Intrusiveness} Preserving subjects' privacy is probably the most important considerations when implementing such a framework in production.
Especially, given the worldwide adoption of the General Data Protection Regulation (GDPR)~\cite{eu:gdpr}.
With respect to this criteria, it is very important to mention that the data collection framework for conducting a user study (experimentation and research) are not the same as for a production system.
Specifically, when conducting an exploratory experiment more data is collected because it is impossible to know in advance what are the exact parameters and data that are required for evaluating the security awareness. 
Therefore, in the experimental framework we collected raw network traffic and sensor data.
This raw data was stored in an anonymized and encrypted form in an isolated environment, and analyzed by authorized users only.
However, in a production environment the collected information can be dramatically minimized: (a) based on the research results and derived insights only the relevant data (that is required for the specific purpose) can be collected and analyzed; (b) analysis of most data can be done on the device and therefore, there is no need to store/transmit raw data out of the device (only aggregated data which does not contain any sensitive information).

Nevertheless, with respect to intrusiveness, the security questionnaire is the least intrusive data sources. 
This is followed by the network traffic monitor and mobile device agent which are more intrusive. 
However, in a production system, where the analysis are done on the device and only aggregated information is stored for a long period; the network traffic monitor and mobile device agent are not more intrusive than most of today’s intrusion detection/prevention systems (IDSs/IPSs), and mobile device management solutions.

%% file: conclusion.tex
\section{\label{sec:conclusion-future-work}Conclusions}

In this paper, we presented a novel framework for evaluating the ISA of smartphone users.
The assessment is based on three data sources, each one of which requires a different level of interaction with the subjects. 
Our results shows that the mobile agent provides the most reliable information; the traffic monitor requires the least interaction with the subjects; and the security questionnaire is the least intrusive among the three.    

We also conducted a comprehensive user study involving 162 smartphone users.
Within this study we used the proposed framework to monitor the behaviour of users while using their own smartphones.
In addition, all subjects were exposed to four simulated challenges that resemble real-life social engineering attacks. using 
Our results indicate that an ISA assessments, which are based on the mobile agent and on the passive traffic monitoring correctly reflect the ability of the subjects to mitigate SE attacks.  
In addition, multiple contradictions were observed between the self-reported behavior of subjects according to the security questionnaire and the actual measured behavior.
These inconsistencies stress the necessity of acquiring objective measurements while assessing ISA of smartphone users.

The practical significance of this study are three-fold.
First, the framework enable organizations to accurately and continuously assess the ISA of their employees without interfering with their everyday tasks.  
Second, the framework enable the early identification of users that are vulnerable to a specific SE attack class, and provide the organization with the exact factors that make a user vulnerable to those attacks. 
Knowing those factors enable an effective, targeted and efficient implementation of countermeasures, such as developing a focused security awareness programs, customize cyber security protection tools to the individual needs of employees, and etc.
Third, because it does not require any interaction with the users, the proposed framework is also scalable and the assessment can be done automatically.

%% file: Resources/Behavior_Questionnaire.tex
\preto\tabular{\setcounter{magicrownumbers}{0}}
\newcounter{magicrownumbers}
\newcommand\rownumber{\stepcounter{magicrownumbers}\arabic{magicrownumbers}}

\newcommand{\crit}[1]{}
\newcommand{\QFour}[1]{\rownumber. & {#1} & \small{$\square$} & \small{$\square$} & \small{$\square$} & \small{$\square$} \\}
\newcommand{\QFive}[1]{\rownumber. & {#1} & \small{$\square$} & \small{$\square$} & \small{$\square$}& \small{$\square$} & \small{$\square$} \\}
\newcommand{\QSix}[1]{\rownumber. & {#1} & \small{$\square$} & \small{$\square$} & \small{$\square$}& \small{$\square$} & \small{$\square$} & \small{$\square$} \\}
\newcommand{\QThree}[1]{\rownumber. & {#1} & \small{$\square$} & \small{$\square$} & \small{$\square$} \\}

\newcolumntype{M}[1]{>{\centering\arraybackslash}m{#1}}

\begin{center}
\scriptsize
  \begin{tabularx}{\linewidth}{p{.01\linewidth} p{0.42\linewidth} M{0.085\linewidth} M{0.085\linewidth} M{0.085\linewidth} M{0.085\linewidth} M{0.085\linewidth}}
	   \multicolumn{2}{p{0.4\linewidth}}{\textbf{What is the likelihood of you to perform the following actions?}} & Never & Unlikely & Medium Likelihood & Very Likely & Always \\ \hline ~\\
       \QFive{Downloaded application from unofficial application store. \crit{[AI1]}}
       \QFive{Install application that requires permissions that are not necessary for its functionality (figure attached). \crit{[AI2]}}
       \QFive{Install applications with a low rating (figure attached). \crit{[AI3]}}
       \QFive{Install application that requires root privileges (figure attached). \crit{[AI4]}}
       \QFive{Approve an application update that requires permissions that are not  necessary for the application’s functionality (figure attached). \crit{[AH1]}}
       \QFive{Verify an application update before approving it. \crit{[AH1]}}
       \QFive{Enter an advertisement while using an application. \crit{[AH2]}}
       \QFive{Enter an advertisement of a lottery. \crit{[AH2]}}
      \QFive{Check which applications are installed on the device (figure attached). \crit{[AH3]}}
      \QFive{Check which applications are running (figure attached). \crit{[AH3]}}
      \QFive{Close applications that are running in the background (figure attached). \crit{[AH3]}}
      \QFive{Delete applications that are not in use. \crit{[AH3]}} 
      \QFive{Enter a website despite a security warning that this site is dangerous (figure attached). \crit{[B1,B5]}} 
      \QFive{Download file from a site that does not use an encryption protocol (figure attached).\crit{[B2]}}
      \QFive{Insert private information in a site which does not use an encryption protocol (figure attached).\crit{[B3]}}
      \QFive{Use your personal password in a site which does not use an encryption protocol (figure attached).\crit{[B3]}}
      \QFive{Enter private information (e.g., phone number, email address) into a pop-up that appears while using an application (figure attached). \crit{[B4]}}
      \QFive{Open an email classified as spam (figure attached). \crit{[VC1]}}
      \QFive{Enter a link sent from unknown party (e.g., via Facebook, WhatsApp, SMS). \crit{[VC2]}}
      \QFive{Download a file sent to you by email from an unknown sender. \crit{[VC2]}}
      \QFive{Use a simple password that contains known personal details (e.g., name, date of birth, phone number).\crit{[A1]}}
      \QFive{Use a password that is constructed from many different digits and characters. \crit{[A1]}}
      \QFive{Use the same password for different services. \crit{[A1]}}
      \QFive{Save your password as plain-text on your device (e.g., in your contacts, document, notes). \crit{[A1]}}
      \QFive{Update your password in case you suspect that someone knows it. \crit{[A1]}}
      \QFive{Update your password at least once a year. \crit{[A1]}}
      \QFive{Use two-factor authentication for email/Facebook (a service in which another authentication is used besides the password - figure attached). \crit{[A2]}}
      \QFive{Use password management services (figure attached). \crit{[A3]}}
      \QFive{Jailbreak/root your device. \crit{[OS2]}}
      \QFive{Use embedded security systems (e.g., firewall, encryption). \crit{[SS1]}}
      \QFive{Install anti-virus application. \crit{[SS2]}}
      \QFive{Ignore a security alert (figure attached). \crit{[SS4]}}
      \QFive{Use screen lock (e.g., PIN code, pattern, fingerprint). \crit{[SS5]}}
      \QFive{Use public (unencrypted) WiFi networks (figure attached).\crit{[N1]}}
      \QFive{Download files while using a public network.\crit{[N2]}}
      \QFive{Use a VPN services while using the internet.\crit{[N3]}}
      \QFive{Use personal services (e.g., bank, online shopping, Facebook) while connected to a public Wi-Fi network \crit{[N4]}}
      \QFive{Auto-connect to Bluetooth devices.\crit{[PC1]}}
      \QFive{Connect your smartphone to foreign device (e.g., a friend’s computer, wireless headphones). \crit{[PC2]}}
     % \QFive{Left your device unattended.\crit{[PC3]}}
\end{tabularx}%
\end{center}

\begin{center}
\scriptsize
\begin{tabularx}{\linewidth}{p{.01\linewidth} p{0.2\linewidth}  M{0.12\linewidth}  M{0.12\linewidth} M{0.12\linewidth} M{0.12\linewidth} M{0.12\linewidth}}
   \multicolumn{2}{p{0.25\linewidth}}{\textbf{How often do you turn off the following components in your device? }}& Always off, unless needed & Turned on when needed, but often forget to turn off  & Turned on if the battery allows & Always on & I didn't know it could be turned off \\ \hline ~\\
    \QFive{GPS  \crit{[PC1]}}
    \QFive{Wi-Fi  \crit{[PC1]}}
    \QFive{NFC \crit{[PC1]}}
    \QFive{Bluetooth  \crit{[PC1]}}
\end{tabularx}%
\end{center}

\begin{center}
\scriptsize
 \begin{tabularx}{\linewidth}{p{.01\linewidth} >{\raggedright\arraybackslash}M{0.41\linewidth} M{0.085\linewidth} M{0.085\linewidth} M{0.085\linewidth} M{0.085\linewidth} M{0.085\linewidth}}
	\multicolumn{2}{p{0.4\linewidth}}{\textbf{How frequently do you update the following services/applications on your smartphone?}} & I use auto-update & When I receive a notification & Once a month & Once in six months & Only when I must \\ \hline ~\\
    \QFive{Operating system. \crit{[OS1]}}
    \QFive{System applications (e.g., Web browser, messaging applications).\crit{[AH1]}}
    \QFive{Security applications (e.g., anti-virus, firewall).\crit{[SS3]}}
    \QFive{Other applications on the device (e.g., Facebook, WhatsApp)\crit{[AH1]}}
\end{tabularx}%
\end{center}
~\\
\newpage

%% file: Resources/Taxonomy_table_with_measurements.tex
{

% new_indexed_criteria = {
% 	'AI1': ['AI1',],
% 	'AI2': ['AI2', 'AH3',],
% 	'AI3': ['AI3',],
% 	'AI4': ['AI4',],
% 	'AI5': ['AI5',],
		
% 	'AH1': ['AH1',],
% 	'AH2': ['AH4', ],
% 	'AH3': ['AH6'],

% 	'B1': ['B1', 'VC1',],
% 	'B2': ['B3'],
% 	'B3': ['B4'],
% 	'B4': ['B5', 'B7'],
% 	'B5': ['B6', 'B8', 'B9'],

% 	'VC1': ['VC2','VC5','VC6'],
% 	'VC2': ['VC4'],

% 	'A1': ['A1', 'A2', 'A3','A4',],
% 	'A2': ['A5'],
% 	'A3': ['new criteria'], 

% 	'OS1': ['OS1'],
% 	'OS2': ['OS3'], 

% 	'SS1': ['SS1',],
% 	'SS2': ['SS2'],
% 	'SS3': ['SS4'],
% 	'SS4': ['SS5'],
% 	'SS5': ['SS7'],

% 	'N1': ['N1','N2'],
% 	'N2': ['N3'],
% 	'N3': ['N4'],
% 	'N4': ['N5'],

% 	'PC1': ['PC1'],
% 	'PC2': ['PC2'],
% 	'PC3': ['PC3'],
% }

\newcolumntype{M}[1]{>{\centering\arraybackslash}m{#1}}
\newcolumntype{G}[1]{>{\raggedright\arraybackslash}m{#1}}
\newcommand{\header}[2]{
 \hline
  \multicolumn{6}{|c|}{\textbf{Focus Area:} #1 \textbf{Sub-Focus Area:} #2} \\
 \hline
}
\newcommand{\criteria}[6]{
   #2 & #1 & #3 & #4 & #5 & #6\\ \hline
}

\newcommand{\criteriatraffic}[4]{
  #1 &  #2 & #3 & #4 & #4 \\ \hline
}

\begin{scriptsize}
\begin{longtable}{|  M{.03\textwidth} | M{.1\textwidth} | G{.18\textwidth} |G{.18\textwidth} | G{.18\textwidth} | G{.1\textwidth} |}

\hline
 \textbf{ID}  & \textbf{Topics} & \textbf{Description} & \textbf{Network Indicators} & \textbf{Mobile Agent Indicators} & \textbf{Questionnaire Indicators} \\ 
\hline
% ######################################################################
\header{Applications}{Application Installation (AI)}
\criteria{Untrusted Sources}
          {AI1}
          {Download applications solely from trusted sources}
          {Monitor accesses untrusted application sources, monitor downloads of APK files.}
          {Check the source of installed applications.}
          {Q1}
\criteria{Permissions}
          {AI2}
          {Does not install applications that require dangerous permissions.}
          {\centering --}
          {Monitor applications that requires dangerous permissions and their Google Play ranking is less than 3.5 \cite{dangerouspermissions}.}          
          {Q2}
\criteria{Rating}
          {AI3}
          {Does not install applications with a low rating.}
          {\centering --}
          {Monitor installed applications with low Google Play rating (less than 3.5).}          
		  {Q3}
\criteria{Root/Jailbreak}
          {AI4}
          {Rarely installs applications that require root privileges.}
          {\centering --}
          {Monitor applications that require root privileges}          
		  {Q4}
% \criteria{Unknown/Unsigned Applications}
%           {AI5}
%           {Does not install unsigned applications, and do not download unknown/unsigned applications.}
%           {Monitor downloads of APK files via unencrypted protocols (i.e., HTTP).}
%           {Check the source of installed applications}               
% 		  {~}
% ######################################################################
\header{Applications}{Application Handling (AH)}
\criteria{Application Updates}
          {AH1}
          {Regularly update applications}
          {\centering --}
          {Check how many of the installed applications are not up-to-date.}  %***
          {Q5, Q6}

% \criteria{Malicious Updates}
%           {AH2}
%           {Verify application update before approving it.}
%           {\centering --}
%           {~}
%     	  {Q13, Q15}	 
% \criteria{Permissions}
%           {AH3}
%           {Does not install applications that require sensitive permissions.}
%           {~}
%           {Monitor applications with low Google play ranking (less than 3.5) that requires dangerous permissions \cite{dangerouspermissions}.}    
%           {Q14}
\criteria{Advertisements}
          {AH2}
          {Rarely clicks on advertisements}
          {Monitor communication with domains associated with monitoring advertisements clicks (e.g., ad.doubleclick.net).} 
          {Monitor communication with domains associated with monitoring advertisements clicks (e.g., ad.doubleclick.net).} 
          {Q7, Q8}
% \criteria{Privacy Settings}
%           {AH5}
%           {Properly configures applications’ privacy settings to preserve privacy.}
%           {\centering --}
%           {~}
%    		  {~}
\criteria{Application Management}
          {AH3}
          {Properly manages running/installed applications}
          {\centering --}
          {Monitor deletion of applications.}
   		  {Q9,Q10,Q11,Q12}          
          
% ######################################################################

\header{Browsing and Communication}{Browser (B)}          
\criteria{Malicious Domain}
          {B1}
          {Does not enter malicious domains}
          {Monitor accesses malicious domains.}
          {Monitor accesses malicious domains.}
		  {Q13}
% \criteria{Certificates – General}
%           {B2}
%           {Does not operate via HTTP in domains that support HTTPS.}
%           {\centering --}
%           {~}   
% 		  {~}
          
\criteria{File Downloads}
          {B2}
          {Prefers to download files via HTTPS.}
          {Monitor downloads of files via HTTP.}   
          {Monitor downloads of files via HTTP.}
          {Q14}
\criteria{Data Confidentiality}
          {B3}
          {Does not send sensitive information via HTTP.}
          {Monitor personal information sent as plaintext (using Regex).}
          {\centering --}
          {Q15, Q16}
\criteria{Pop-ups and Advertisements}
          {B4}
          {Does not insert private information into popups.}
          {Monitor transmission of private information around the time at which a pop-up was clicked.}
          {\centering --}
		  {Q17}	
\criteria{Certificates}
          {B5}
          {Operates in accordance with system/browser alerts and does not use untrusted certificates.}
          {Monitor sessions that include untrusted certificates that were not terminated immediately after the handshake.}
          {Monitor accesses to domains with untrusted certificates}
		  {Q13}
% \criteria{Advertisements/ Advertising Sites}
%           {B7}
%           {Does not insert private information on advertisers' sites.}
%           {Monitor transmission of private information around the time in which a pop-up was clicked.}
%           {Monitor transmission of private information around the time in which a pop-up was clicked.}
% 		  {~}
% \criteria{Certificates – Installation}
%           {B8}
%           {Operates in accordance with system/browser alerts and does not install untrusted certificates.}
%           {Monitor sessions that include untrusted certificates that were not terminated immediately after the handshake.}
%           {Monitor sessions that include untrusted certificates that were not terminated immediately after the handshake.}
% 		  {~}
% \criteria{Certificates – Validation}
%           {B9}
%           {Validates domain certificates and operates in accordance with system/browser alerts.}
%           {Monitor sessions that include untrusted certificates that were not terminated immediately after the handshake.}
%           {Monitor sessions that include untrusted certificates that were not terminated immediately after the handshake.}
%           {Q8}
          
% \criteria{Plugins, Add-ons, Extensions}
%           {B10}
%           {Properly manages the plugins/add-ons/extensions installed on his/her browser and erases unwanted plugins}
%           {\centering --}
%           {~}
%           {~} 

% ######################################################################
\header{Browsing and Communication}{Virtual Communication (VC)}         
% \criteria{Malicious Hyperlinks}
%           {VC1}
%           {Does not click on suspicious hyperlinks.}
%           {Monitor accesses to malicious domains.}
%           {Monitor accesses to malicious domains.}
%   		  {Q6}
          
\criteria{Spam}
          {VC1}
          {Does not open emails/messages received from unknown senders (or spam).}
          {Monitor communication with spam domains around the time of communicating with email services.}
          {Monitor opens spam emails.}
          {Q18}
          
% \criteria{Unknown Sender/Spam – Hyperlinks}
%           {VC3}
%           {Validates URLs embedded in emails.}
%           {\centering --}
%           {~}
%           {Q6}
\criteria{Malicious Attachments/URL}
          {VC2}
          {Does not download/execute attachments, or click on URL's received from unknown senders.}
          {Monitor communication with malicious domains around the time of communicating with email services.}
          {Monitor clicks on links sent via SMS from unknown senders.}
          {Q19, Q20}
% \criteria{Advertisements}
%           {VC5}
%           {Does not open advertisement emails received from unknown senders.}
%           {\centering --}  
%           {Monitor requests to URLs that received via SMS from unknown sender.}
%           {~}
% \criteria{Junk Mailbox}
%           {VC6}
%           {Does not open/read messages/emails classified as spam.}
%           {\centering --} 
%           {Monitor opened spam emails.}
%           {~}
% ######################################################################
\header{Browsing and Communication}{Account (A)}

\criteria{Password Management}
          {A1}
          {Updates passwords regularly, use unguessable and diverse passwords. Does not store passwords unsafely}
          {\centering --}  
          {\centering --}
          {Q21, Q22, Q23, Q24, Q25, Q26}

% \criteria{Passwords – Updates}
%           {A1}
%           {Updates passwords regularly and does not use default passwords.}
%           {\centering --}  
%           {\centering --}
%           {Q40, Q41}
          
% \criteria{Passwords – Diversity}
%           {A2}
%           {Does not use the same password for different services.}
%           {\centering --}  
%           {\centering --} 
%           {Q38}
% \criteria{Passwords – Strength}
%           {A3}
%           {Uses strong and unguessable passwords.}
%           {\centering --}  
%           {\centering --} 
%           {Q36, Q37}
% \criteria{Passwords – Storage}
%           {A4}
%           {Does not store passwords unsafely.}
%           {\centering --}  
%           {\centering --}     
%           {Q39}
\criteria{Two Factor Authentication}
          {A2}
          {Uses two-factor authentication mechanisms.}
          {\centering --}  
          {Monitor SMS with authentication codes (i.e., monitor messages associated with two-factor authentication mechanisms).}    
          {Q27}
% \criteria{Privacy Settings}
%           {A6}
%           {Restricts the privacy settings in social media.}
%           {\centering --}  
%           {\centering --}  
%           {~}

\criteria{Keychain Services}
          {A3}
          {Uses password management services.}
          {Monitor accesses domains associated with password management services.}  
          {Check for the existence of password management applications.} % TODO: add examples
          {Q28}

% ######################################################################
\header{Device}{Operating systems (OS)}
\criteria{Updates}
          {OS1}
          {Uses an updated OS and configures auto-updates for critical services}
          {Monitor the device model and OS version (from HTTP user-agent string) and verify that this version is the latest.}  
          {Verify that the OS version is the latest.} 
          {Q45}
% \criteria{Vulnerable OS}
%           {OS2}
%           {Does not use an OS with known vulnerabilities and stays updated regarding known vulnerabilities of his/her OS.}
%           {\centering --}  
%           {-}  
%           {~}
\criteria{Root/Jailbreak}
          {OS2}
          {Does not root or jailbreak the device.}
          {\centering --}  
          {Check whether the device is rooted.}  
          {Q29}
% ###################################################################### 
% \header{Device}{Data Privacy (DP)}
% \criteria{Encryption}
%           {DP1}
%           {Encrypts the drive of the device.}
%           {\centering --}  
%           {-}  
%           {~}
% \criteria{Backup}
%           {DP2}
%           {Uses backup services.}
%           {\centering --}  % ?
%           {-}  
%           {~}
% \criteria{Private Data}
%           {DP3}
%           {Does not store sensitive material on the device.}
%           {\centering --}  
%           {-}  
%           {~}
% ######################################################################          

\header{Device}{Security Systems (SS)}
\criteria{Embedded Security Systems} % TODO delete!!
          {SS1}
          {Uses embedded security systems.}
          {\centering --} 
          {\centering --} 
          {Q30}
\criteria{Anti-virus}
          {SS2}
          {Uses anti-virus application regularly to scan the device.}
          {Monitor accesses domains associated with anti-virus services}  
          {Check for the existence of anti-virus applications} 
          {Q31}
% \criteria{Mobile Device Management (MDM)}
%           {SS3}
%           {Uses MDM services.}
%           {\centering --}  
%           {-}
%           {~}
\criteria{Security Updates}
          {SS3}
          {Updates security systems.}
          {Monitor accesses to domains associated with anti-virus services}  
          {Monitor that security applications are up-to-date}  
  		  {Q47}
\criteria{Security Alerts}
          {SS4}
          {Operates in accordance with security alerts (i.e., does not ignore security alerts).}
          {\centering --}  
          {\centering --}    
          {Q32}
% \criteria{Remote Deletion Services}
%           {SS6}
%           {Installs remote deletion services on the device.}
%           {\centering --}  
%           {-}
%           {~}
\criteria{Lock Screen}
          {SS5}
          {Uses PIN-code/pattern/ fingerprint.}
          {\centering --}  
          {Check whether a lock screen PIN-code/pattern is defined.} 
          {Q33}
% ########################################################## Complete measurement:
\header{Communication Channels}{Networks (N)}
% \criteria{Wi-Fi – Auto-connect}
%           {N1}
%           {Does not auto-connect to unknown Wi-Fi networks.}
%           {\centering --}  
%           {Does not connect to unencrypted networks.} % ????
%           {Q30}
\criteria{Wi-Fi - Unencrypted Networks}
          {N1}
          {Does not connect to unencrypted networks.}
          {\centering --}  
          {Monitor connections to unencrypted Wi-Fi networks.}
          {Q34}
\criteria{Wi-Fi - Unencrypted Networks}
          {N2}
          {Does not download files on unencrypted Wi-Fi networks}
          {\centering --}  
          {Monitor files downloaded while connected to unencrypted Wi-Fi networks.}
          {Q35}
\criteria{VPN}
          {N3}
          {Uses VPN services on public networks.}
          {\centering --}  
          {Monitor VPN client applications installed to the device.}  
          {Q36}
\criteria{Wi-Fi - Unencrypted Networks}
          {N4}
          {Does not transmit private data via unencrypted channels.}
          {\centering --}  
          {Monitor private data transmitted while connected to unencrypted Wi-Fi networks}  
		  {Q37}
% ####################################################
\header{Communication Channels}{Physical Channels (PC)}
\criteria{Connectivity}
          {PC1}
          {Enables Bluetooth, Wi-Fi, NFC, and GPS while they are in use and only connects trusted Bluetooth and NFC devices, and trusted Wi-Fi networks.}
          {\centering --}  
          {Check whether the Bluetooth and/or NFC are turned on while no device is connected. Check whether the GPS is always turned on.} %?
          {Q38, Q41, Q42, Q43, Q44 }
\criteria{USB/Charging Connections}
          {PC2}
          {Does not use a USB to connect the device to unknown devices.}
          {\centering --}  
          {-}    
          {Q39}
% \criteria{Physical access}
%           {PC3}
%           {Does not keep the device unattended and uses a PIN-code.}
%           {\centering --}  
%           {Check weather a lock screen PIN-code/pattern is defined.}                
% 		  {Q1}
\end{longtable}
\end{scriptsize}

}